\documentclass[%
aip,
amsmath,amssymb,
reprint,
nofootinbib%
]{revtex4-2}
\usepackage{nomencl}
\makenomenclature
\renewcommand\nomgroup[1]{%
  \item[\bfseries
  \ifstrequal{#1}{A}{Nomenclature}{%
  \ifstrequal{#1}{N}{Abbreviations}{%
  \ifstrequal{#1}{B}{Subscripts}{%
  \ifstrequal{#1}{C}{Superscripts}{%
  \ifstrequal{#1}{D}{Greek Letters}{%
  \ifstrequal{#1}{X}{Other symbols}{}}}}}}%
]}

\usepackage{graphicx,array,booktabs,rotating,parskip}
\usepackage{dcolumn}
\usepackage{bm}
\usepackage{mathptmx}
\usepackage{comment}
\usepackage{enumitem}
\usepackage{afterpage,float,color,xcolor}
\usepackage{multirow}
\usepackage{lipsum} 
\usepackage{hyperref}
\usepackage{etoolbox} 
\usepackage{tabularx,tabulary}
\usepackage[capitalize]{cleveref}
\usepackage{times}

\hypersetup{
	colorlinks,
	linkcolor={blue!100!black},
	citecolor={blue!100!black},
	urlcolor={blue!100!black}
}
\newcommand{\PRLsep}{\noindent\makebox[\linewidth]{\resizebox{0.3333\linewidth}{1pt}{$\bullet$}}\bigskip}
\makeatletter
\newcommand\footnoteref[1]{\protected@xdef\@thefnmark{\ref{#1}}\@footnotemark}
\makeatother

\newcommand{\etal}{\textit{et al.}}
\usepackage[page]{totalcount}
\usepackage{etoolbox,fancyhdr,xcolor}%
\usepackage{ragged2e}
\usepackage[displaymath,pagewise]{lineno}
\begin{document}

\preprint{AIP/123-QED}

\title[\textbf{Journal to be decided} (2026) $|$ Manuscript under preparation]{Transonic flow past the complex cavity-sub-cavity configurations}
\author{Anagha K}
\email{kuniyilanagha@gmail.com}
\affiliation{Department of Mechanical and Aerospace Engineering,\\ 
Indian Institute of Technology Hyderabad, Telangana – 502284, India}

\author{Harshit Bansal}
\email{ae22d200@smail.iitm.ac.in}
\affiliation{Department of Aerospace Engineering,\\ 
Indian Institute of Technology Madras, Chennai, Tamil Nadu – 600036, India}

\author{Jaysinh J. Patel}
\email{jaysinhpatel.drdl@gov.in}
\affiliation{Defence Research and Development Laboratory (DRDL),\\
Defence Research \& Development Organization (DRDO), Telangana – 500058, India}

\author{Rajesh Kumar}
\email{rajeshk.drdl@gov.in}
\affiliation{Defence Research and Development Laboratory (DRDL),\\
Defence Research \& Development Organization (DRDO), Telangana – 500058, India}

\author{R. Sriram}
\email{r.sriram@iitm.ac.in}
\affiliation{Department of Aerospace Engineering,\\ 
Indian Institute of Technology Madras, Chennai, Tamil Nadu – 600036, India}

\author{Gnanaprakash K}
\email{gnan@mae.iith.ac.in}
\affiliation{Department of Mechanical and Aerospace Engineering,\\ 
Indian Institute of Technology Hyderabad, Telangana – 502284, India}

\author{Niranjan S. Ghaisas}
\email{nghaisas@mae.iith.ac.in}
\affiliation{Department of Mechanical and Aerospace Engineering,\\ 
Indian Institute of Technology Hyderabad, Telangana – 502284, India}

\author{Hideaki Ogawa}
\email{hideaki.ogawa@aero.kyushu-u.ac.jp}
\altaffiliation{(also at Japan Aerospace Exploration Agency, JAXA).}
\affiliation{Department of Aeronautics and Astronautics,\\ 
Kyushu University, Fukuoka – 8190001, Japan}

\author{S. K. Karthick}
\email[Corresponding author: ]{skkarthick@mae.iith.ac.in}
\affiliation{Department of Mechanical and Aerospace Engineering,\\ 
Indian Institute of Technology Hyderabad, Telangana – 502284, India}

\date{\today}

\begin{abstract}

The study investigates the physics of unsteady flow in complex cavity geometries operating in the transonic regime. A two-dimensional Detached Eddy Simulation (DES) approach is used for the preliminary analysis. The cavity configuration examined in this work arises from the integration of a scramjet engine with a launch vehicle. In this integrated geometry, the isolator section serves as a deep sub-cavity, while the Single Expansion Ramp Nozzle (SERN) constitutes the primary cavity. The combined arrangement therefore constitutes a complex cavity-sub-cavity system, which is referred to as such throughout the paper. The qualitative analysis revealed a feedback loop within the complex cavity-sub-cavity system, leading to high-pressure oscillations across the geometry. A monotonic increase in pressure loading is observed with increasing Mach number. Varying the cavity topology demonstrated that modifications to the primary cavity geometry strongly alter shear-layer dynamics and significantly affect the pressure distribution within the cavity-sub-cavity system. To mitigate adverse pressure oscillations, passive control strategies, including trailing-edge wall chamfering and a ventilated (slotted) sub-cavity, are investigated. Among the configurations studied, the slotted sub-cavity case exhibits the most pronounced suppression of pressure loads, particularly on the sub-cavity end wall. Spectral Proper Orthogonal Decomposition (SPOD) analysis also revealed the restructuring of dominant coherent modes in response to topological variations and to the implementation of passive control, providing insight into the underlying governing mechanism.

\end{abstract}

\maketitle
\nomenclature[A]{$M$}{Mach number (-)}
\nomenclature[A]{$T$}{Static temperature (K)}
\nomenclature[A]{$p$}{Static pressure (Pa)}
\nomenclature[A]{$\rho$}{Density (kg/m$^3$)}
\nomenclature[A]{$L$}{Cavity length (m)}
\nomenclature[A]{$D$}{Cavity depth (m)}
\nomenclature[A]{$l$}{Sub-cavity length (m)}
\nomenclature[A]{$d$}{Sub-cavity depth (m)}
\nomenclature[A]{$\theta$}{Nozzle angle (deg)}
\nomenclature[A]{$x$}{Streamwise coordinate (m)}
\nomenclature[A]{$y$}{Wall-normal coordinate (m)}
\nomenclature[A]{$t$}{Time (s)}
\nomenclature[A]{$u$}{$x$-component of velocity (m/s)}
\nomenclature[A]{$v$}{$y$-component of velocity (m/s)}
\nomenclature[A]{$E$}{Total energy per unit mass (J/kg)}
\nomenclature[A]{$e$}{Internal energy per unit mass (J/kg)}
\nomenclature[A]{$f$}{Frequency (Hz)}
\nomenclature[A]{$s$}{Surface coordinate (m)}
\nomenclature[A]{$\Pi$}{Normalized power spectra  $\left[fG_{xx}(f)/p_r^2\right]$ (-)}
\nomenclature[A]{$\hat{q}(\mathbf{x},\omega)$}{Fourier-transformed flow-field fluctuations}
\nomenclature[A]{$\hat{Q}_{\omega}$}{Frequency-domain data matrix}
\nomenclature[A]{$\mu$}{Dynamic viscosity (kg/(m·s))}
\nomenclature[A]{$\nu$}{Kinematic viscosity (m$^2$/s)}
\nomenclature[A]{$\tau$}{Viscous shear stress (Pa)}
\nomenclature[A]{$\omega$}{Vorticity (1/s)}
\nomenclature[A]{$S_{ij}$}{Strain-rate tensor}
\nomenclature[A]{$\delta_{ij}$}{Kronecker delta (-)}
\nomenclature[A]{$\Delta$}{Difference or step size}
\nomenclature[A]{$\Lambda$}{Eigenvalues (-)}
\nomenclature[A]{$\Psi$}{Eigenvectors (-)}
\nomenclature[A]{$\Phi$}{Spatial modes (-)}
\nomenclature[A]{$m$}{Spatial mode number (-)}
\nomenclature[A]{$n$}{Rossiter mode number (-)}
\nomenclature[A]{$Re$}{Reynolds number (-)}
\nomenclature[A]{$St$}{Strouhal number (-)}
\nomenclature[A]{$\kappa$}{Convective velocity ratio ($U_c/U_\infty$)}
\nomenclature[A]{$\dot{m}$}{Non-dimensional mass flow rate}
\nomenclature[A]{$\eta$}{Mass flow ratio}
\nomenclature[A]{$k$}{Turbulent kinetic energy (J/kg)}
\nomenclature[A]{$R$}{Turbulent viscosity ratio}
\nomenclature[A]{$I$}{Turbulent intensity}
\nomenclature[A]{$C$}{Model constant (-)}

\nomenclature[B]{$i,j,k$}{Spatial indices}
\nomenclature[B]{$0$}{Stagnation quantity}
\nomenclature[B]{$\infty$}{Freestream quantity}
\nomenclature[B]{$r$}{Reference quantity}
\nomenclature[B]{$1,2,3$}{Probe locations}

\nomenclature[C]{$\overline{\Box}$}{Time-averaged quantity}
\nomenclature[C]{$\Box'$}{Fluctuating quantity}
\nomenclature[C]{$\Box_\sigma$}{Standard deviation}

\nomenclature[N]{CFD}{Computational Fluid Dynamics}
\nomenclature[N]{RANS}{Reynolds-Averaged Navier-Stokes}
\nomenclature[N]{URANS}{Unsteady Reynolds-Averaged Navier-Stokes}
\nomenclature[N]{LES}{Large Eddy Simulation}
\nomenclature[N]{DES}{Detached Eddy Simulation}
\nomenclature[N]{PSD}{Power Spectral Density}
\nomenclature[N]{FFT}{Fast Fourier Transform}
\nomenclature[N]{SPOD}{Spectral Proper Orthogonal Decomposition}
\nomenclature[N]{BG}{Baseline Geometry}
\nomenclature[N]{SG}{Single Expansion Ramp Nozzle Geometry}
\nomenclature[N]{IG}{Inverted Single Expansion Ramp Nozzle Geometry}
\printnomenclature

\section{Introduction} \label{sec:Introducation}
Cavity flow is often defined as a complex flow problem occurring in high-speed aerodynamic applications~\cite{East1966,Unalmis2004,CisnerosGaribay2022,Loupy2018,Rokita2012,Zhao2019,Oza2015,McCloud2011}. The interaction of the freestream with the relatively stagnant fluid inside the cavity forms a separated shear layer~\cite{Brandeis1982,Murray2001} at the leading edge which becomes susceptible to Kelvin-Helmholtz (K-H) instability, leading to the roll-up of coherent vortical structures. Impingement of these vortices on the cavity's trailing edge generates strong acoustic and pressure disturbances. These disturbances propagate upstream and interact with the shear layer at the leading edge, thereby reinforcing the instability mechanism. This hydrodynamic-acoustic coupling~\cite{rossiter1964} establishes a feedback loop that results in self-sustained oscillations~\cite{rossiter1964,Yokoyama2009,ROWLEY2002,Lee2008} of the shear layer at certain flow conditions. 

While such self-sustained oscillatory behavior can be advantageous in many applications, such as enhancing mixing and providing stable recirculation zones for flameholding~\cite{Gruber2004,Zhu2025,BenYakar1998} in high-speed propulsion systems, it is often accompanied by significant adverse effects, including high-amplitude pressure fluctuations~\cite{Zhang1990,CHARWAT1961,CHARWAT1961-II,HELLER1975,Tam1978}, increased aerodynamic drag~\cite{gharib1987}, and severe aeroacoustic noise~\cite{Chandra2005,Hardin1977}. The prevalence of cavity flows across multiple aerospace systems, such as weapon bays~\cite{Loupy2018,Barakos2014}, landing gear wells~\cite{Oza2015_CavityFlow_LG}, missile launch compartments~\cite{saravanan2009}, and scramjet combustor modules~\cite {CisnerosGaribay2022,CAO20233147}, has motivated decades of fundamental and applied research activities. 

Cavities are often classified based on their length ($L$) to depth ($D$) ratio into three categories~\cite{CHARWAT1961,CHARWAT1961-II}: a. open cavity~\cite{Murray2009} ($L/D < 10$), b. transitional cavity~\cite{Lada2010} ($10 < L/D < 14$), and c. closed cavity~\cite{Chung2003} ($L/D > 14$). In open cavities, the separated shear layer spans directly from the leading to the trailing edge without reattaching to the cavity floor, resulting in a strong hydrodynamic-acoustic feedback mechanism. Transitional cavities display intermediate behavior, characterized by progressive thickening of the shear layer and partial reattachment within the cavity floor. In closed cavities, the shear layer reattaches to the cavity floor upstream of the trailing edge, thereby forming two distinct separation bubbles within the cavity. Moreover, such an event suppresses coherent vortex impingement and weakens the classical feedback loop.

Extensive studies on open-type~\cite{rossiter1964,heller1971,gharib1987,ROWLEY2002,colonius1999,FRANKE1975} cavity flows have been conducted over several decades to investigate their self-sustained oscillatory characteristics. These studies have led to well-established scaling laws~\cite{rossiter1964,heller1971} and modal descriptions~\cite{Doshi2022} that form the foundation of our current understanding of cavity flow dynamics. Cavity flows have been extensively investigated, from the empirical characterization of discrete tonal oscillations and the formulation of Rossiter’s semi-empirical frequency model~\cite{rossiter1964} to the mechanistic elucidation of the hydrodynamic–acoustic feedback process, as systematically formalized by Rockwell and Naudascher~\cite{Rockwell1978}. Even now, they remain one of the most thoroughly studied canonical problems in high-speed flows. Subsequent investigations on these flows highlighted the importance of three-dimensional effects~\cite{Chung2001,Maull1963,Woo2008}, nonlinear mode interactions~\cite{schmidt2022}, and compressibility influences~\cite{Beresh2016}. Within this broader context, numerous studies have further demonstrated that cavity-flow dynamics are strongly influenced by key governing parameters, including Mach number~\cite{Jia2024,Bhaduri2025}, Reynolds number~\cite{Bhaduri2025}, cavity geometry~\cite{Casper2016}, and the incoming boundary-layer characteristics~\cite{Savchenko2020}.

The available literature on open-type~\cite{} cavity flows spans a broad range of Mach numbers, extending from subsonic~\cite{Murray2001,Gloerfelt2003} ($M_\infty<1$) to supersonic~\cite{Panigrahi2019,karthick2021,Li2013} ($M_\infty>1$) regimes, with the governing dynamics transitioning from predominantly acoustic feedback in subsonic flows to coupled shock-shear-layer interactions and compressibility-dominated acoustic feedback mechanisms in supersonic conditions. The transonic regime ($0.8 \leq M_\infty \leq 1.2$) remains a comparatively less-explored area despite its practical significance, arising from high-subsonic Mach number flight or breaking the sonic barrier during supersonic dashes. The intrinsic complexity of this regime, marked by the simultaneous presence of compressibility effects, emerging shock structures, and evolving hydrodynamic-acoustic coupling, makes the systematic isolation and characterization of defining events challenging.

A range of active~\cite{colonius1999,Mongeau1998,Micheau2004} and passive control strategies~\cite{Lad2018,Lawson2009,Bhaduri2025}  has been investigated to mitigate the dominant pressure oscillations in unsteady cavity flows. Researchers like  Lad \etal~\cite{Lad2018} experimentally demonstrated that the introduction of a secondary recess or a sub-cavity within a supersonic cavity can modify shear-layer evolution and alter dominant resonance characteristics. Building on this premise, subsequent investigations~\cite{Bhaduri2025} into sub-cavity effects in open cavities, including front- and aft-wall sub-cavity configurations, have firmly established sub-cavity geometry as a robust and effective passive control parameter for modulating oscillatory behavior in compressible cavity flows. Preliminary research conducted by some of the authors \cite{CherishmaNSSW2024,Cherishma2024,Cherishma2025,AnaghaNSSW2024,Kuniyil2025} of the present paper on a rectangular cavity with a sub-cavity at a transonic Mach number describes the sustenance of unsteady flow. 

Research works of Cherishma \etal~\cite{CherishmaNSSW2024} indicated the crucial role of sub-cavity aspect ratio in the unsteadiness developed within the cavity system. The researchers classified the sub-cavity system into two extremes: shallow, with a low aspect ratio (length-to-depth ratio, $l/d$), and deep sub-cavity. The deep sub-cavity system produced noticeable unsteadiness, whereas the shallow case showed reduced unsteadiness compared with the baseline cavity without a sub-cavity. The mean pressure at the sub-cavity end wall increased by 14.8\% relative to the baseline case. Furthermore, the power spectral density (PSD) obtained from the fast fourier transform (FFT) analysis indicated a 97\% increase in spectral energy, highlighting significantly stronger pressure fluctuations within the sub-cavity region as the sub-cavity aspect ratio ($l/d$) increases. Moreover, the researchers have adopted a specific passive control method \cite{Cherishma2024} by chamfering the sharp corners in the main cavity to reduce shear-layer unsteadiness in a deep-cavity-sub-cavity system. The considered control technique reduced the peak spectral power of the pressure fluctuations at the shear-layer midpoint by 83\% at a freestream Mach number of $M_\infty=1.2$. Another passive control approach involving a serpentine-shaped sub-cavity geometry was also investigated by Cherishma \etal~\cite{Cherishma2025}. The modified geometry reduced pressure fluctuations at the sub-cavity end wall and suppressed the peak spectral power by 90\% at a freestream Mach number of $M_\infty = 0.9$.

Joshi~\etal~\cite{Joshi2025} investigated the application of passive control techniques for a simple rectangular cavity with deep sub-cavity configuration at a supersonic freestream Mach number of $M_\infty = 2.5$. Geometric modifications, such as chamfering sharp edges, and trailing-edge modifications, such as step-up (increasing trailing-edge height) and step-down (decreasing trailing-edge height), were highlighted. The results indicated that the chamfered configuration significantly amplified pressure fluctuations at the sub-cavity end-wall and cavity floor by almost two times that of the freestream pressure ($p_\infty$). Configurations like step-up showed an increase in pressure fluctuations at the sub-cavity end-wall by 22\% and a decrease by 8-23\% along the cavity floor, whereas step-down configuration showed an increase in pressure fluctuations at the sub-cavity end-wall by 33\% and along the cavity floor by 58-113\%.

\begin{figure*}
    \centering
    \includegraphics[width=1\linewidth] {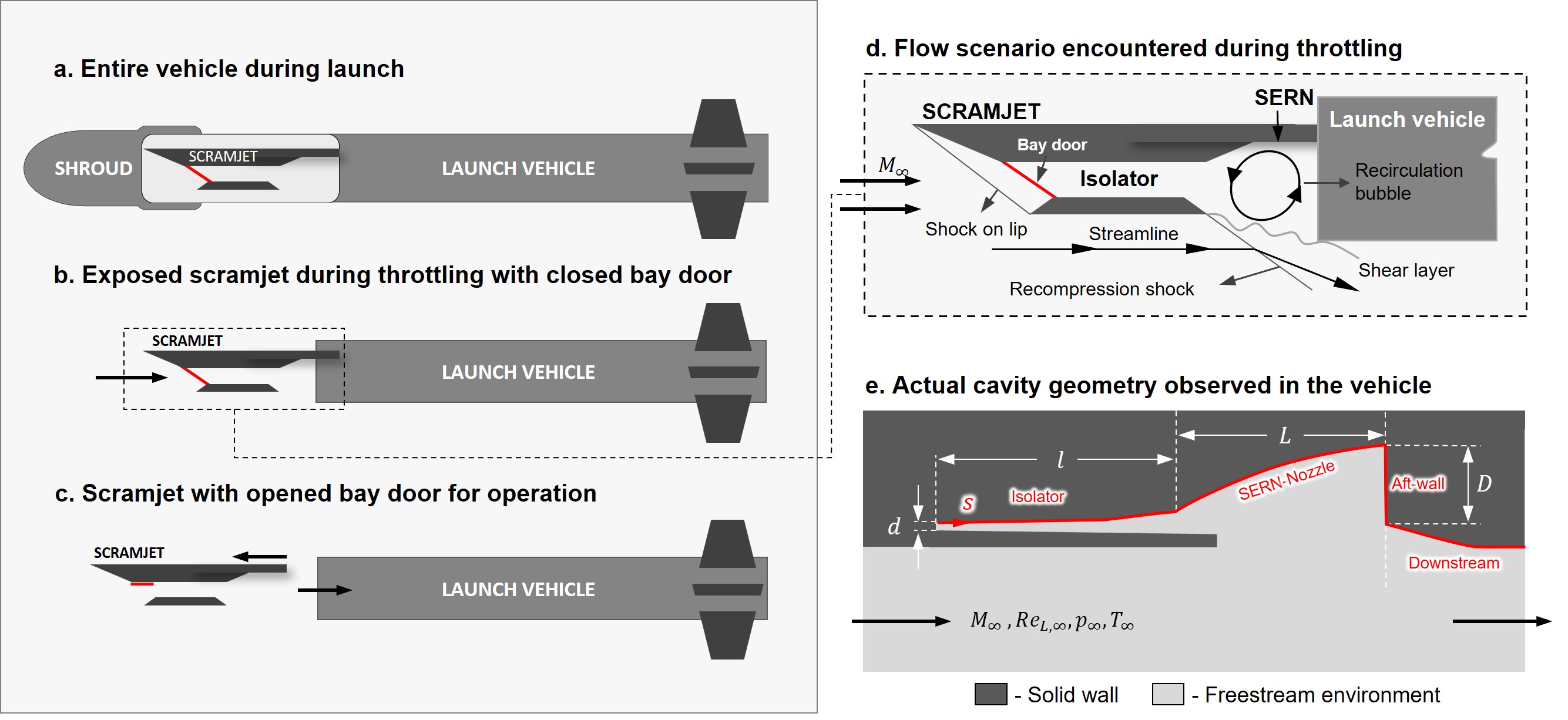}
    \caption{Schematic illustrating different stages of scramjet engine deployment using a launch vehicle: (a) launch stage—the complete launch vehicle with the scramjet engine enclosed within a shroud; (b) throttling stage—the scramjet engine exposed to the freestream with the bay door closed; (c) detachment stage—the launch vehicle releasing the scramjet with the bay door open; (d) schematic of the flow scenario during the throttling stage, highlighting key aerodynamic features; and (e) schematic of the approximated two-dimensional complex cavity-sub-cavity geometry employed in the present study, representing the actual configuration encountered during the real time flight in the throttling stage. The surface coordinate along the cavity is shown and denoted as $s$. The regions of isolator, SERN nozzle, aft wall, and downstream of the cavity are marked in red. The cavity and sub-cavity length ($L$ and $l$) and depth ($D$ and $d$) are also marked in the schematic for reference.}
    \label{fig:problem_statement2}
\end{figure*}

Although extensive research has been conducted on open cavities with deep-sub-cavity systems across various regimes, the majority of existing studies have focused on simplified geometries~\cite{Lad2018,CherishmaNSSW2024,Cherishma2025,Joshi2025} that rarely represent the complex cavity-sub-cavity systems encountered in realistic aerospace systems. However, it is worth noting that limited research is available on the standalone complex cavity shapes~\cite{Casper2014,Ozalp2010,Luo2011}.  In any case, practical configurations, such as engine-airframe integrations, bay-door assemblies, or scramjet vehicle interfaces, cavity geometries often deviate significantly from regular shapes due to manufacturing, structural, aerodynamic, and propulsion constraints. These irregular or compound cavities often involve multiple interacting sub-cavities with varying aspect ratios, which fundamentally alter the hydrodynamic–acoustic coupling mechanisms. 

Kuniyil \etal~\cite{AnaghaNSSW2024} illustrated the intricate flow physics involved in a complex cavity-sub-cavity configuration encountered in a certain aerospace system. They highlighted the major similarities and noticeable differences between the regular and complex cavity-sub-cavity systems when deploying the passive control technique, such as chamfering. In a more detailed study of Kuniyil \etal~\cite{Kuniyil2025}, they also highlighted the influence of curvature on the main rectangular cavity-sub-cavity configuration previously studied by Cherishma \etal~\cite{CherishmaNSSW2024,Cherishma2024,Cherishma2025}. They transformed the rectangular cavity into a complex cavity with a deep sub-cavity system. One notable observation was that the introduction of concave curvature significantly amplifies pressure oscillations and peak pressure loads within the cavity. The mean pressure load at the sub-cavity end-wall was observed to reach up to 1.5 times the freestream pressure. Furthermore, spectral analysis using the Fast Fourier Transform (FFT) indicated that the cavity configuration with concave curvature exhibited approximately 60\% higher peak spectral power than the rectangular cavity-sub-cavity configuration, implying stronger unsteadiness.

All these previous investigations remain largely preliminary and limited in comprehensively addressing the complex flow physics associated with realistic cavity-sub-cavity geometries. The present work seeks to address this gap by investigating the flow dynamics over a specific complex cavity-sub-cavity configuration arising from the structural integration of a scramjet module with a launch vehicle. Although three-dimensional (3-D) effects can influence detailed flow structures, previous studies\cite{ROWLEY2002,Panigrahi2019,karthick2021} have shown that the primary unsteady dynamics and dominant resonance modes in cavity flows are adequately captured using two-dimensional (2-D) approximations. Hence, the rest of the discussion focuses on the 2-D numerical analysis, which also provides a computationally efficient framework for resolving unsteady flow interactions and proposes suitable control methods for attenuation. 

A multi-level cavity system comprising a main cavity corresponding to the SERN (Single Expansion Ramp Nozzle) nozzle and an embedded sub-cavity corresponding to the scramjet isolator region is shown in Figure~\ref{fig:problem_statement2}. The coupling between these two cavities introduces unsteady flow behavior similar to that observed in conventional geometries, thereby forming the complex cavity-sub-cavity system. These systems encounter different operating conditions in reality. One such operation is considered here. During launch operations, a certain scramjet-launch vehicle system experiences a series of dynamic exposure conditions. The shroud enclosing the scramjet is first ejected while throttling (Figure~\ref{fig:problem_statement2}-a), exposing the engine directly to the freestream (Figure~\ref{fig:problem_statement2}-b). The scramjet module subsequently separates from the launch vehicle under operational flow conditions (Figure~\ref{fig:problem_statement2}-c). Prior to detachment, the high-speed flow impinging on the complex cavity-sub-cavity system within the scramjet-vehicle configuration generates a separated shear layer, a recompression shock near the leading edge, and a recirculation zone in the SERN region, as illustrated in Figure~\ref{fig:problem_statement2}-d. These flow structures closely resemble those observed in conventional cavity flows and, collectively, give rise to complex hydrodynamic-acoustic interactions that can significantly affect vehicle stability, aerodynamic control, and engine operability.

\begin{figure*}
    \centering
   \includegraphics[width=1\linewidth] {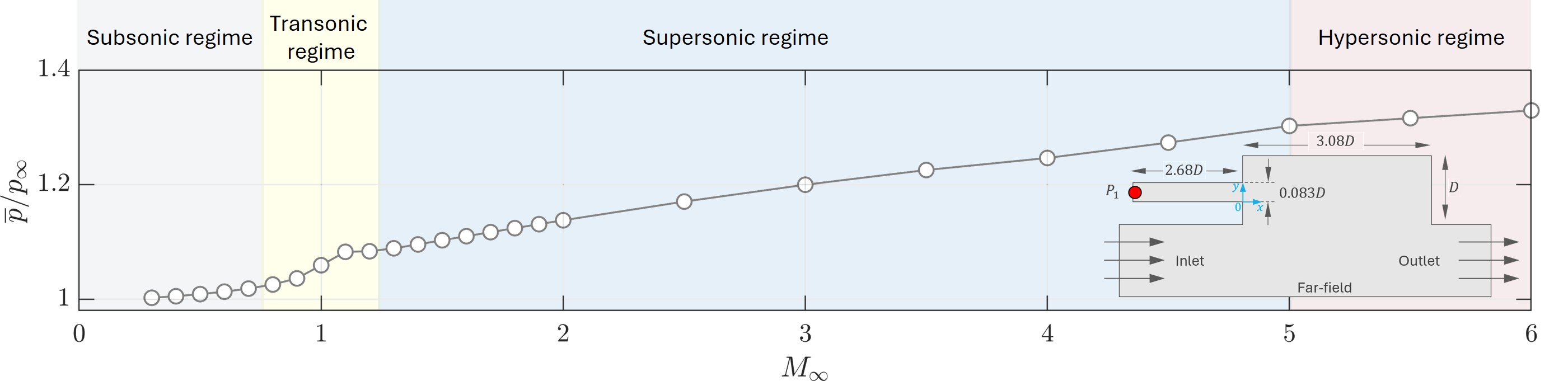}
    \caption{Variation of the non-dimensionalized time-averaged static pressure ($\overline{p}/p_\infty$) observed at the sub-cavity end wall (probe $P_1$) for a range of Mach numbers ($M_\infty$) for a simple cavity-sub-cavity system from the RANS based simulation. The inset in the bottom right indicates the respective geometric configuration of the simple cavity-sub-cavity system exposed to the freestream conditions (Note: non-dimensionalizing variable used is $p_\infty=$ 2.6kPa.}
    \label{fig:P_vs_M}
\end{figure*}

A preliminary numerical investigation, using the steady-state RANS (Reynolds Averaged Navier-Stokes) equations to obtain the time-averaged flow behavior, is conducted for a simple cavity with a sub-cavity system at different Mach numbers. Figure~\ref{fig:P_vs_M} shows the variation of mean pressure ($\overline{p}/p_\infty$) at the sub-cavity end-wall (probe $P_1$) over a range of Mach numbers ($M_\infty$). The results indicate that the mean pressure load at the probe $P_1$ increases with Mach number in the subsonic regime, reaches a peak within the transonic range, and subsequently decreases. Beyond the transonic regime, at higher Mach numbers, the mean pressure increases monotonically. This non-monotonic behavior observed in the transonic regime reflects the coexistence of localized subsonic and supersonic regions within the flowfield. Such coupled and competing mechanisms highlight the intrinsic complexity of transonic cavity-sub-cavity flows, making this regime both challenging and essential to investigate, as undertaken in the present study.

\begin{figure}
    \centering
    \includegraphics[width=1\linewidth] {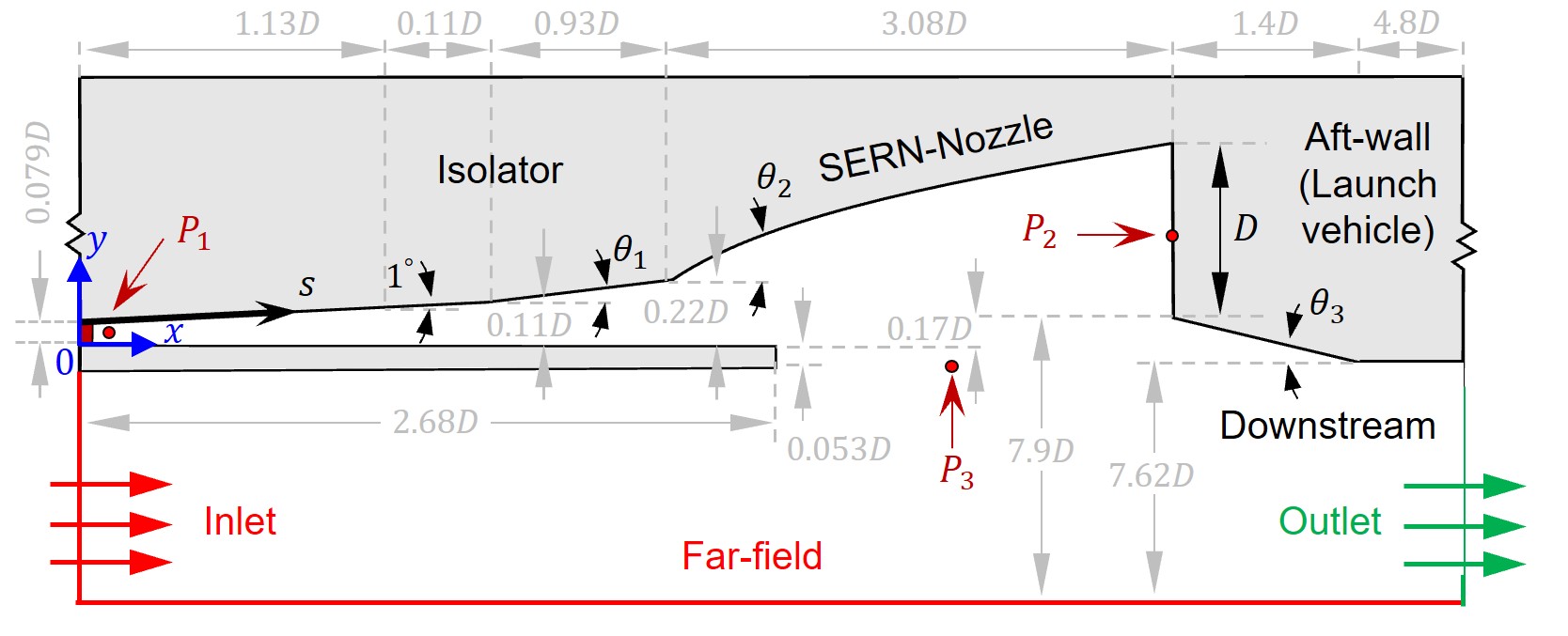}
    \caption{Schematic shows the computational domain of the complex cavity-sub-cavity system encountered in the scramjet-launch vehicle interface with specified boundary conditions (inlet, far-field, and outlet). Geometrical parameters are non-dimensionalized using the cavity depth $D$. The surface coordinate along the cavity is shown and denoted as $s$. Probing points $P_1$, $P_2$, and $P_3$ (at locations $[x_1/D,y_1/D] =$ [0, 0.06], $[x_2/D,y_2/D] =$[5.76, 1.22], and $[x_3/D,y_3/D] =$ [4.22, 0.01], respectively) are indicated for flow diagnostics. The origin is located at the cavity leading edge (highlighted in blue). The flow is from left to right (Note: The schematic is not drawn to scale).}
    \label{fig:computational_domain}
\end{figure}

The paper is organized as follows. Section \ref{sec: geo_det} provides details of the cavity-sub-cavity configuration investigated in this study. The numerical methodology is outlined in Section \ref{sec: numerical methodology}, covering the computational domain and the flow solver. Section \ref{sec: gov eqn} covers the governing equations involved in the numerical study. Section~\ref{sec: independence} covers grid and time-step independence studies, as well as solver validation against experimental data. Section \ref{sec: results} presents the results and discussions, encompassing unsteady statistics, the flow physics of the complex cavity-sub-cavity system, and the dominant spectral modes governing the flow. Finally, Section \ref{sec:conclusion} summarizes the study's key findings.

\section{Geometrical Details} \label{sec: geo_det}
Figure~\ref{fig:computational_domain} shows the two-dimensional computational domain for the complex cavity-sub-cavity configuration resulting from the attachment of the scramjet engine to the launch vehicle, as discussed in Section \ref{sec:Introducation}. The geometry is parameterized using the primary cavity depth `$D$'. The cavity is of the open type with a length ($L$) to depth ($D$) ratio of $[L/D] = 3.08$. A regime spanning $[x/D]=4$ upstream and $[x/D]=6.2$ downstream of the cavity is considered for the flow domain. The domain is extended transversely $[y/D] = 7.9$ to prevent shock reflections and backflow at the boundaries. The flow is from left to right, with the origin positioned at the cavity's leading edge (marked in blue). Any point along the cavity surface can be referenced using the surface length coordinate $s$ (see Figure~\ref{fig:computational_domain}). The sub-cavity region features a wall of depth `$d$', and the sub-cavity depth varies along the axial direction. The total length of the sub-cavity is $[l/D] = 2.68$ and features dual divergent sections located at $[x/D] = 1.24$ and $[x/D] = 2.17$, with divergent angles of $\theta_1 = 5^\circ$ and $\theta_2 = 25.4^\circ$,  respectively. The primary cavity replicates a Single Expansion Ramp Nozzle (SERN) shape extending from the sub-cavity region. A ramp with an inclination angle of $ \theta_3 =11.3^\circ$ is also located at the trailing edge of the primary cavity. Unsteady statistics are collected at three probing locations $P_1$, $P_2$, and $P_3$  (see Figure~\ref{fig:computational_domain}), which are later utilized to understand the flow physics and also to compare with the experimental results. The simulations are carried out for freestream conditions corresponding to an altitude of 25 km above the sea level, with a freestream pressure $p_\infty = 2607.7$ Pa and a freestream temperature $T_\infty=216.67$ K over a range of Mach numbers ($M_\infty=0.9, 1.0, 1.1$, and $1.2$). More detailed flow parameters considered for the numerical simulation are listed in Table \ref{tab:freestream_params}.

\begin{table}
\caption{\label{tab:freestream_params}Details of the flow and geometrical parameters utilized in the numerical simulations for different freestream Mach numbers.}
\begin{ruledtabular}
\begin{tabular}{lc}
\textbf{Parameter} & \textbf{Value} \\
\midrule
Freestream static pressure ($p_\infty$, Pa) & 2607.7 \\
Freestream static temperature ($T_\infty$, K) & 216.7 \\
Freestream velocity ($u_\infty$, m/s) & 265-354 \\
Freestream Mach number ($M_\infty$, -) & 0.9-1.2 \\
Reference length ($L$, m) & 11.7 \\
Reynolds number ($Re_{\infty,L}\times 10^6$, -) & $9.1$-$12.1$ \\
\end{tabular}
\end{ruledtabular}
\end{table}

\section{Numerical Methodology}
\label{sec: numerical methodology}
A commercial flow solver from ANSYS-Fluent\textsuperscript{\tiny\textregistered} package\cite{ANSYS2013_FluentTheory} is used to perform the numerical simulations. The solver has been extensively used to study various compressible flow phenomena, including cavity flows\cite{Pey2014_RampCavityFlow}, shear-shock interactions\cite{VamsiKrishna2024_LeadingEdgeBluntness}, and shock-train dynamics\cite{Gugulothu2020_ScramjetShockTrain, Gnani2018_ShockTrainStructure}, collectively demonstrating its capability and accuracy in resolving high-speed flows. 
A two-dimensional numerical approach is a good starting point for understanding the global flow physics governing such complex phenomena. They are computationally cost-effective and offer flexibility in grid generation as well, and are thereby used widely for studying such complex flows~\cite{Sun2018_SpanwiseEffects_RectangularCavity,Liu2024_SupersonicCavityFlows,Fedorova2012_JetInjectionCavity,RoshanSah2021_SupersonicTurbulentCavity,Liu2024_CavitatingFlowHydrofoil}. In the present study, a two-dimensional framework serves as a preliminary investigation to understand the large-scale phenomena observed in complex cavity-sub-cavity systems. A Detached Eddy Simulation (DES) model was employed to investigate the unsteady characteristics of the flow observed over the complex cavity-sub-cavity configuration shown in Figure~\ref{fig:computational_domain}. The DES method was chosen over conventional Unsteady RANS (URANS) and high-fidelity LES (Large Eddy Simulation) for the following reasons: a. URANS has inherent limitations to capture flows dominated by global instabilities; b. LES requires highly refined near-wall grid resolution, resulting in high computational cost; c. DES offers a balanced trade-off by resolving large-scale, unsteady turbulent structures in massively separated flows such as cavity flows~\cite{karthick2021}, while maintaining comparatively lower computational costs; d. DES effectively captures the dominant longitudinal flow modes, which are sufficient to resolve shock–shear layer interactions~\cite{karthick2021}, thereby providing an effective means to meet the objectives of our study.

A detailed two-dimensional computational domain with boundary conditions used for the simulation is shown in Figure~\ref{fig:computational_domain}. The inlet employs a pressure far-field boundary condition, while the outlet is specified as a pressure outlet. Air is used as the working fluid and is modeled as an ideal gas. Fluid viscosity is calculated using Sutherland’s three-coefficient model. The inlet turbulence is specified through turbulent intensity ($I=5\%$) and turbulent viscosity ratio ($R_t = 10$).
Near-wall turbulence is resolved with the RANS-based $k-\omega$ SST (Shear Stress Transport) along with the DDES (delayed-DES) shielding function with default constants given in the solver. The surrounding solid walls are kept at adiabatic conditions. The wall-turbulence parameter is maintained at $y^+ < 1$ near solid surfaces to ensure accurate boundary-layer resolution. A coupled pressure-based solver with a compressibility correction is preferred over a traditional density-based solver for high-speed flows. The coupled pressure-based solver is shown to reasonably well predict the supersonic or hypersonic flow field, including a large region of low $Re$ flow, across the domain~\cite {karthick2021}.
 Spatial gradients are discretized using the Green-Gauss node-based method and second-order upwind schemes. A bounded second-order implicit scheme with iterative time advancement is applied for transient formulations. Initially, the steady-state solution is achieved via hybrid initialization before switching to the transient mode. Hybrid initialization is favored again for rapid solution
convergence. It predicts the values in the fluid domain almost closer to the expected through internal iterations (typically in 10-20 iterations) by
solving a simplified system of flow equations (see Section~\ref{sec: gov eqn}). The analysis is conducted using a developed flow over a total duration of $[t/T]= 250$, where $T = D/u_\infty$ and $u_\infty$ is the freestream inflow velocity.

\begin{figure*}
    \centering
    \includegraphics[width=1\linewidth] {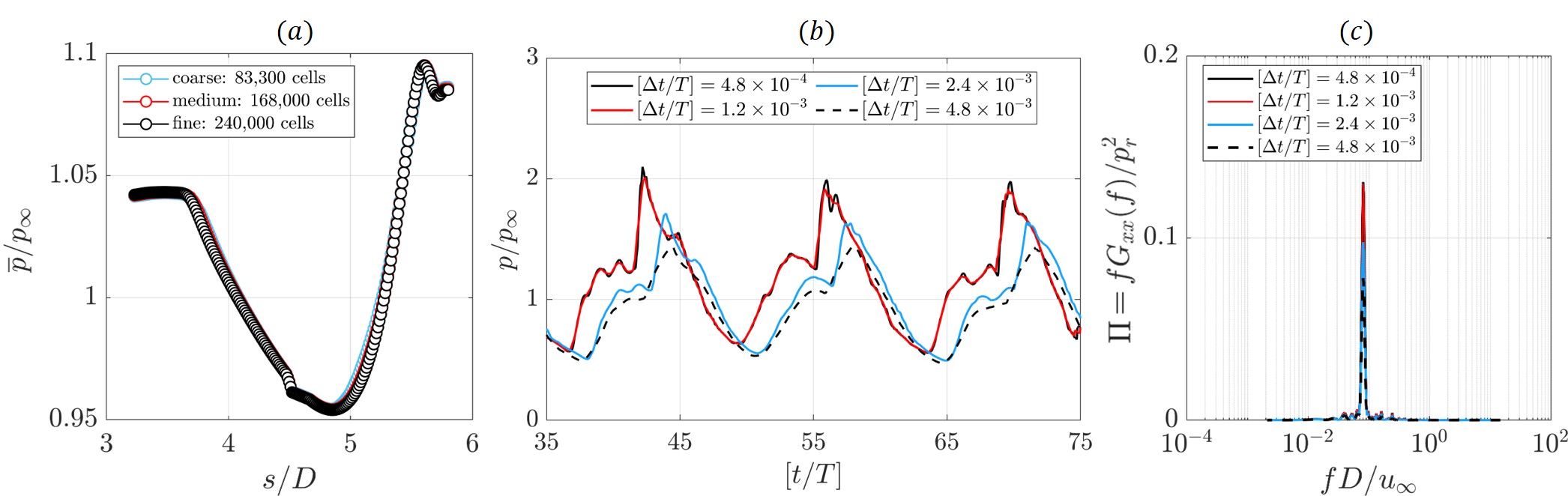}
    \caption{(a) Normalized time-averaged static pressure ($\overline{p}/p_\infty$) variation along the cavity floor ($s/D$, see Figure~\ref{fig:computational_domain} for definition) is plotted to assess grid independence using three grid densities: coarse - 83,300 cells; medium - 168,000 cells; and fine - 240,000 cells. (b) Non-dimensionalized temporal variation of normalized static pressure ($p/p_\infty$) monitored at shear-layer midpoint (probe location $P_3$, see Figure~\ref{fig:computational_domain}) for four different non-dimensionalized time step sizes ($\Delta t /T$) towards time-step independence study: $4.8\times 10^{-3}$, $2.4\times 10^{-3}$, $1.2\times 10^{-3}$, and $4.8\times 10^{-4}$. (c) Non-dimensionalized Power Spectral Density ($\Pi = fG_{xx}/p_r^2$) plotted against non-dimensionalized frequency ($fD/u_\infty$) for the static pressure obtained at probe location $P_3$ for different time-step sizes. The non-dimensionalizing variables used here are $p_r$ = 1 kPa, $p_\infty$ = 2.6 kPa, $D= 3.8$ m, and $u_\infty=354.08$ m/$s$. Flow is simulated at a freestream Mach number of $M_\infty=1.2$.}
    \label{fig:independence study}
\end{figure*}

\begin{table}
\caption{\label{tab:grid_details}Details of the computational grids used in the grid independence study are summarized below. 
Here, $L_x$ and $L_y$ indicate the total number of cells in the streamwise and transverse directions across the entire domain, respectively, whereas $C_x$ and $C_y$ specify the number of cells within the cavity-sub-cavity region along the streamwise and transverse directions, respectively.}
\begin{ruledtabular}
\begin{tabular}{lccc}
\textbf{Grids} & \textbf{Total Cells} & $L_x \times L_y$ & \textbf{$C_x \times C_y$} \\
\midrule
coarse & 83,300  & 490 $\times$ 170 & 150 $\times$ 120 \\
medium & 168,000 & 600 $\times$ 280 & 200 $\times$ 170 \\
fine & 240,000 & 800 $\times$ 300 & 250 $\times$ 200 
\end{tabular}
\end{ruledtabular}
\end{table}

\section{Governing Equations}
\label{sec: gov eqn}
The basic fluid flow Navier–Stokes equations, including continuity, momentum,
and energy, are solved numerically. The governing conservative form of the equations is averaged using the principles of Favre-Averaging (Favre-averaged Navier–Stokes equations) as shown in the following equations in the index notation, and the indices increment represents the three flow axes ($x$, $y$, $z$):

\begin{equation}
\frac{\partial \rho}{\partial t} + \frac{\partial}{\partial x_i} (\rho u_i) = 0, 
\label{eq:continuity}
\end{equation}

\begin{equation}
\frac{\partial}{\partial t} (\rho u_i) + \frac{\partial}{\partial x_j} (\rho u_i u_j) = -\frac{\partial p}{\partial x_i} + \frac{\partial \tau^t_{ij}}{\partial x_j}, 
\label{eq:2}
\end{equation}

\begin{equation}
\frac{\partial}{\partial t} (\rho E_0) + \frac{\partial}{\partial x_j} (\rho E_0 u_j) = -\frac{\partial}{\partial x_j} (p u_j - q^t_j + u_i \tau^t_{ij}). 
\label{eq:energy}
\end{equation}


\begin{equation}
- \rho \overline{u_i' u_j'} =
\mu_t \left( \frac{\partial \overline{u_i}}{\partial x_j}
+ \frac{\partial \overline{u_j}}{\partial x_i} \right)
- \frac{2}{3} \rho k \, \delta_{ij}
\label{eq:boussinesq}
\end{equation}

\begin{equation}
\tau_{ij}^t = \left[ (\mu + \mu_t) \left( \frac{\partial u_i}{\partial x_j} + \frac{\partial u_j}{\partial x_i} - \frac{2}{3} \frac{\partial u_k}{\partial x_k} \delta_{ij} \right) \right] - \frac{2}{3} \rho k \delta_{ij},
\label{eq:tau}
\end{equation}

\begin{equation}
q_j^{t}
=
- \left(
\frac{\mu}{Pr}
+
\frac{\mu_t}{Pr_t}
\right)
\frac{\partial h}{\partial x_j}
\label{eq:q_t}
\end{equation}

Bossinique's hypothesis~\ref{eq:boussinesq} is used in Eqs.~\ref{eq:continuity}-\ref{eq:energy} to resolve the terms such as the viscous stress ($\tau_{ij}^{t}$) and heat flux ($q_j^{t}$), as given in Eqs.~\ref{eq:tau} and~\ref{eq:q_t}, where the subscript `$t$' represents the terms arising due to turbulence. The unknown turbulent quantities are resolved using the two-equation k-$\omega$ SST (Shear Stress Transport) model~\cite{Acquaye2016_EvaluationTurbulence,Acquaye2016_ValidationWrayAgarwal,Moura2015_BoundaryLayerShock}, in which the turbulent kinetic energy 
$k$ and the specific dissipation rate $\omega$ are obtained from the 
following transport equations:

\begin{equation}
\frac{\partial (\rho k)}{\partial t}
+\frac{\partial (\rho k u_i)}{\partial x_i}
=P_k-\beta^* \rho k \omega+\frac{\partial}{\partial x_i}\left[(\mu + \sigma_k \mu_t)\frac{\partial k}{\partial x_i}
\right]
\label{eq:k}
\end{equation}

\begin{equation}
\frac{\partial (\rho \omega)}{\partial t}
+
\frac{\partial (\rho \omega u_i)}{\partial x_i}
=
\alpha \frac{\omega}{k} P_k
-
\beta \rho \omega^2
+
\frac{\partial}{\partial x_i}
\left[
(\mu + \sigma_\omega \mu_t)
\frac{\partial \omega}{\partial x_i}
\right]
\label{eq:omega}
\end{equation}

In Eqs.~(\ref{eq:k}) and~(\ref{eq:omega}), $P_k$ represents the production 
of turbulent kinetic energy, $\beta^*$ and $\beta$ are model constants, 
and $\sigma_k$ and $\sigma_\omega$ denote the turbulent Prandtl numbers 
for $k$ and $\omega$, respectively. The turbulent (eddy) viscosity is computed as

\begin{equation}
\mu_t = \frac{\rho k}{\max(\omega,\, F_2 S)}
\end{equation}
where $S = \sqrt{2 S_{ij} S_{ij}}$ is the strain-rate magnitude and $F_2$ is a blending function. In the two-dimensional DES framework, turbulent structures in detached flow regions away from walls are modeled using LES-like subgrid-scale models, while wall-bounded turbulence is addressed using traditional RANS formulations. The transition between RANS and LES behavior is governed by a length-scale criterion. The turbulence length scale is defined as 

\begin{equation}
l_{DES} = \min \left( l_{RANS},\, C_{DES}\Delta \right),
\label{eq:ldes}
\end{equation}

 where $l_{RANS}$ is the modeled RANS length scale. For two-equation models such as $k$-$\omega$ SST, the $l_{RANS}$ can be expressed as
 \begin{equation} 
 l_{RANS} = \frac{\sqrt{k}}{\beta^* \, \omega}
 \end{equation}

 $\Delta$ in equation~(\ref{eq:ldes}) is the local grid spacing (often taken as the maximum of $\Delta x, \Delta y$), and $C_{DES}$ is a model constant ($\simeq 0.65$). In the DES formulation, the RANS length scale is replaced by $l_{DES}$ in the destruction term of the turbulent kinetic energy equation~(\ref{eq:k}). Accordingly, the dissipation term is modified as

\begin{equation}
\beta^* \rho k \omega \;\rightarrow\; \beta^* \rho k \frac{\sqrt{k}}{l_{DES}}.
\end{equation} 
 
 When $l_{RANS} < C_{DES}\Delta$, the model defaults to the RANS mode, which is appropriate in near-wall attached boundary layers. Conversely, when $l_{RANS} > C_{DES}\Delta$, the model switches to LES-like behavior, enabling the resolution of large turbulent structures in separated flow regions away from walls. More detailed explanations of the solver schemes and turbulence closure models are available in the Fluent\textsuperscript{\tiny\textregistered} technical guide~\cite{ANSYS2013_FluentTheory}.
 

\section{Independence studies and solver validation}
\label{sec: independence}
\subsection{Grid independence study}
A two-dimensional planar computational domain, as illustrated in Figure~\ref{fig:computational_domain}, is used for the simulation. The computational mesh is generated using ANSYS ICEM\textsuperscript{\tiny\textregistered}~\cite{ANSYS2023_ICEM}, with the domain discretized into structured quadrilateral cells. Over 94\% of the cells exhibit an equisize skewness~\cite{ANSYS2023_ICEM} parameter around 0.24, indicating high mesh quality. The wall turbulence parameter is maintained at $y^+ < 1$ throughout the domain to ensure accurate boundary-layer progression\cite{ANSYS2023_ICEM} in both the streamwise and transverse directions, and is restricted to a maximum factor of 1.1, particularly near the cavity, to capture shock-shear interactions accurately. Three grid densities were considered to evaluate grid independence using steady RANS simulation: a. coarse grid (0.083 million cells); b. medium grid (0.168 million cells); and c. fine grid (0.24 million cells). The cell density variation for each grid is detailed in Table~\ref{tab:grid_details}. The steady-state time-averaged wall-static pressure ($\overline{p}/p_\infty$) is monitored along the primary cavity floor for all grid types and plotted in Figure~\ref{fig:independence study}a. The computed $\overline{p}/p_\infty$ values show good convergence among the grids. The maximum deviation between the medium and fine grid is approximately 1.48\% (observed at $s/D = 4.72$). Consequently, the fine grid was chosen for all subsequent simulations, as it provides sufficient resolution to capture shear–shock interactions accurately without further mesh refinement.

\begin{figure*}
    \centering
    \includegraphics[width=0.79\linewidth] {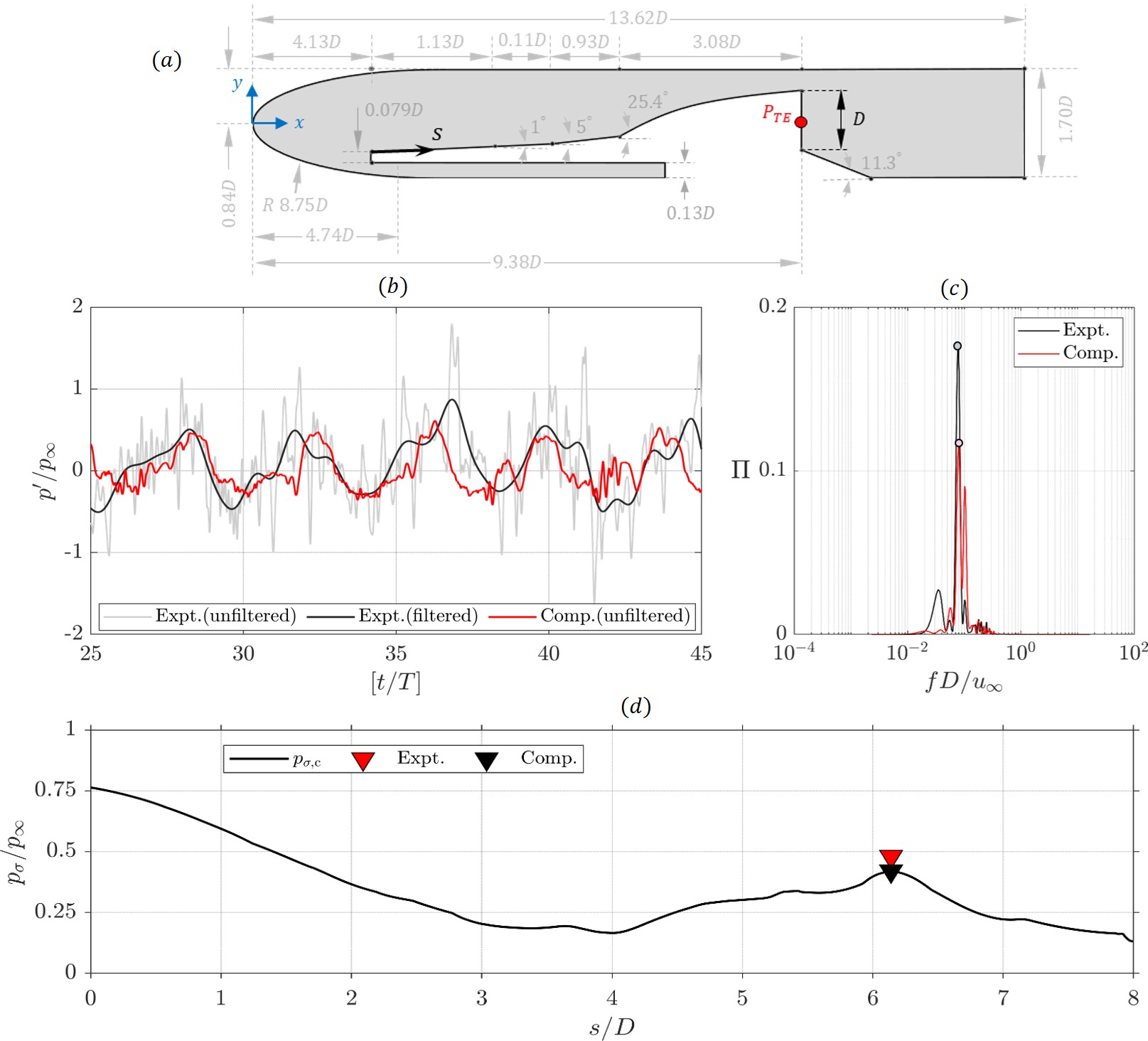}
    \caption{ (a) Two-dimensional schematic of the experimental model in the central plane ($z/D=0$) with various geometrical details. The model geometry is non-dimensionalized using cavity depth $D$. Pressure probe at the trailing-edge location $P_{TE}$ is indicated for reference. The surface coordinate along the cavity is shown and denoted as $s$, originating from the sub-cavity end-wall. (b) Non-dimensional fluctuating pressure ($p'/p_\infty$) measured at probe $P_{TE}$ ($s/D= 6.12$) obtained from numerical computation and experimental measurements. The gray shaded line represents the unfiltered experimental signal, the black line denotes the filtered experimental data, and the red line indicates the unfiltered numerical data; (c) The non-dimensional Power Spectral Density ($\Pi = fG_{xx}/p_r^2$) is plotted against the non-dimensional frequency ($fD/u_\infty$) for the fluctuating pressure signal at the same probe location; (d) The standard deviation of the fluctuating pressure for both the experimental and numerical datasets is shown, where the black line denotes the RMS (Root Mean Square) value of static pressure obtained from the numerical computations across the cavity surface. The non-dimensionalizing variables used here are $p_r$ = 1 kPa, $p_\infty$ = 2.6 kPa, $D= 3.8$ m, and $u_\infty=354.08$ m/$s$. Flow is simulated at a freestream Mach number of $M_\infty=0.9$.}
    \label{fig:experiment_validation}
\end{figure*}

\begin{table}
\caption{\label{tab:exp_details}Details of the experimental flow conditions and model's geometrical parameters observed in the perforated transonic induction-type wind tunnel experiments.}
\begin{ruledtabular}
\begin{tabular}{lc}
\textbf{Parameter} & \textbf{Value} \\
\midrule
Test section size ($L\times W\times H$, m) & $0.61 \times 0.1 \times 0.2$ \\
Model cavity depth ($D$, m) &  0.014\\
Total pressure ($p_0$, kPa) & 101.1 \\
Total temperature ($T_0$, K) & 300 \\
Freestream kinematic viscosity ($\nu_\infty$  $\times 10^{-5}$ , m$^2$/s) & 1.17 \\
Freestream velocity ($u_\infty$, m/s) & 290\\
Mach Number ($M_\infty$, - ) & 0.9 \\
Reynolds number ($Re_\infty/m$, $ \times 10^7$ m$^{-1}$) & 2.5\\
\end{tabular}
\end{ruledtabular}
\end{table}

\subsection{Time-step independence study}

A time step independence study is performed on the complex cavity-sub-cavity system using the fine grid for the geometry shown in Figure~\ref{fig:computational_domain} to ensure adequate temporal resolution. Four non-dimensionalized time steps ($\Delta t/T$) are considered: a. $4.8\times 10^{-3}$; b. $2.4\times 10^{-3}$; c. $1.2\times 10^{-3}$; and d. $4.8\times 10^{-4}$. The time step sizes are chosen based on the physical requirements to resolve the dominant frequency, guidance from values reported in the literature, and considerations of computational resource constraints. The static pressure ($p/p_\infty$) at the probe $P_3$ (shear-layer midpoint) is monitored, with at least three dominant cycles captured at each time step for comparison, as shown in Figure~\ref{fig:independence study}b. The results obtained show that the finest time-step ($\Delta t/T = 4.8\times 10^{-4}$) and the moderately fine time-step ($\Delta t/T = 1.2\times 10^{-3}$) are nearly identical, with their profiles almost overlapping. In contrast, the other two time-steps produce distinctly different trends. The moderately fine time-step ($\Delta t/T = 1.2\times 10^{-3}$) is adopted for the remainder of the numerical simulations, as it accurately captures the flow dynamics while significantly reducing computational cost. The wall-static pressure signal ($p/p_\infty$), containing 20 dominant unsteady cycles at probe $P_3$ is obtained, and its power spectral density (PSD) is determined by performing a fast fourier transform (FFT) as shown in Figure~\ref{fig:independence study}c. The spectrum plot reveals a dominant peak at a Strouhal number of $St\approx0.08$, for all the time-step sizes considered. The collapse of the dominant frequency peak indicates that the chosen temporal resolutions are sufficient to accurately capture the primary unsteady dynamics of the complex-cavity flow. This further confirms that the selected moderately fine time-step ($\Delta t/T= 1.2\times 10^{-3}$) provides adequate temporal resolution to resolve the unsteady flow phenomena, including the primary oscillations and associated harmonics, without introducing significant numerical damping or aliasing. Accordingly, this time step is deemed suitable for all subsequent simulations.

\subsection{Solver validation} \label{ssec:solv_valid}
Figure~\ref{fig:experiment_validation} shows the computational results validated against experimental measurements at a specific possible transonic freestream Mach number of $M_\infty=0.9$, based on the limitation of the available experimental facility. The experiments were conducted in the transonic flow facility at the Gas Dynamics Laboratory, Indian Institute of Technology Madras (IITM), India. The detailed experimental conditions and model parameters employed in the present study are summarized in Table~\ref{tab:exp_details}. The experiments were performed in the perforated transonic induction tunnel, which operated via an ejector mechanism. Jet entrainment downstream of the test section creates a vacuum that draws ambient air into the jet. The test section has a volume of 610 mm (length) $\times$ 100 mm (span) $\times$ 200 mm (height). The test model schematic as shown in Figure~\ref{fig:experiment_validation}a was mounted between the tunnel sidewalls and examined under transonic flow conditions ($T_0 = 300$ K, and $p_0 = 101$ kPa). The upper surface of the model consists of an elliptic forebody to avoid a weak leading-edge shock\cite{Das2017}, followed by a complex cavity-sub-cavity system, a downstream ramp (to represent the launch body), and a flat afterbody, as illustrated in Figure~\ref{fig:experiment_validation}a. The model was fabricated using an FDM (Fused Deposition Modeling) based 3D printer with a layer resolution of 80 $\mu$m and can accommodate pressure transducers. The lateral extensions serve as supports for wall mounting, while the model's central section is exposed to the flow. 

The pressure measurement is obtained at trailing-edge wall location (see Figure~\ref{fig:experiment_validation}a, probe $P_{TE}$). Time-resolved unsteady pressure data were recorded using two Endevco\textsuperscript{\tiny\textregistered} 8530C-15-120 transducers at a sampling rate of 250 kHz, with 1 million samples acquired. Figure~\ref{fig:experiment_validation}b-d shows the results from numerical computations using a moderate time step $[\Delta t/T] = 1.2\times 10^{-3}$ on a fine grid (240,000 cells), validated against experiments at $M_\infty = 0.9$. Figure~\ref{fig:experiment_validation}b shows the non-dimensionalized fluctuating pressure ($p'/p_\infty$) at the probe $P_{2}$ ($s/D = 6.12$, see Figure~\ref{fig:computational_domain}), compared with experimental measurements at the same location (see Figure~\ref{fig:experiment_validation}b, probe $P_{TE}$). The gray and black lines represent the unfiltered and filtered experimental fluctuating pressure, respectively. The computational results closely agree with the filtered experimental data, reproducing its temporal trends fairly well. Furthermore, PSD determined by performing FFT analysis of both pressure measurements indicates that their non-dimensionalized spectral peak frequencies coincide at $fD/u_\infty \sim 0.08$ as shown in Figure~\ref{fig:experiment_validation}c. Figure~\ref{fig:experiment_validation}d shows that the standard deviation of non-dimensionalized pressure between computational and experimental results differs by 3.2\%, further confirming that the unsteady computation, performed with the chosen moderate time-step size, fairly captures the mean and spectral content of the unsteady flow.

\begin{figure*}
    \centering
    \includegraphics[width=1\linewidth] {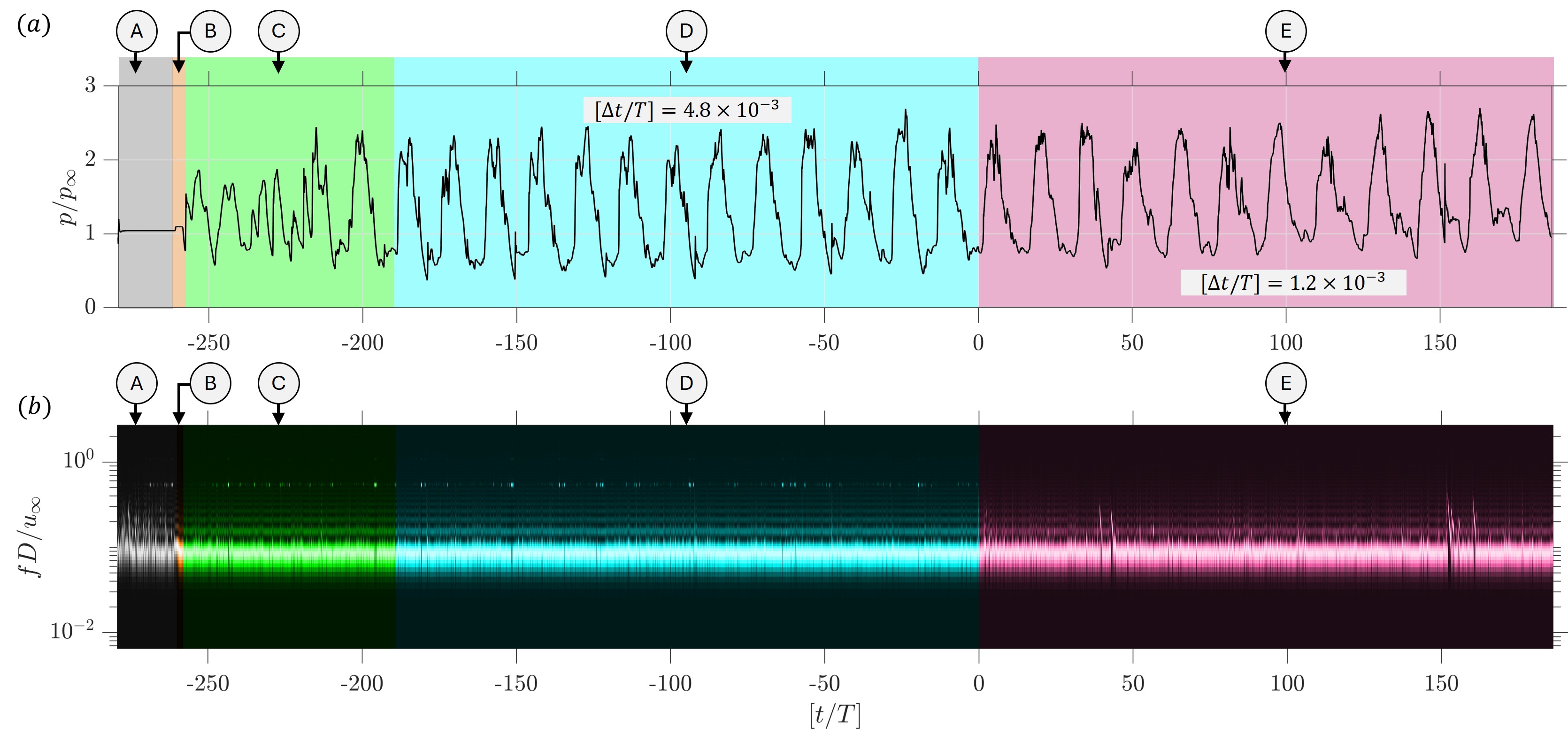}
    \caption{(a) Non-dimensionalized static pressure ($p/p_{\infty}$) variation as a function of non-dimesnionalized time ($t/T$) obtained at the sub-cavity end-wall ($s/D=0$) in flow of $M_\infty=1.2$. The plot illustrates different stages of the unsteady simulation: A. Pseudo–time stepping through the RANS initialization; B. Application of excitation to initiate DES; C. Forced transition phase under coarse time stepping; D. Development of sustained oscillations with refined (fine) time stepping; and E. Established periodic behavior in the statistically stationary DES regime. The colored regions (gray, green, blue, and pink) correspond to these respective phases of the simulation; (b) Locally normalized spectrogram of the pressure signal, where the power at each frequency is scaled between 0 and 1. (Note: non-dimensionalizing variables used here are $p_\infty=$ 2.6kPa, $D=3.8$ m, and $u_\infty = 354.08$ m/$s$).}
    \label{fig:time-stepping}
\end{figure*}

\begin{figure*}
    \centering
    \includegraphics[width=0.9\linewidth] {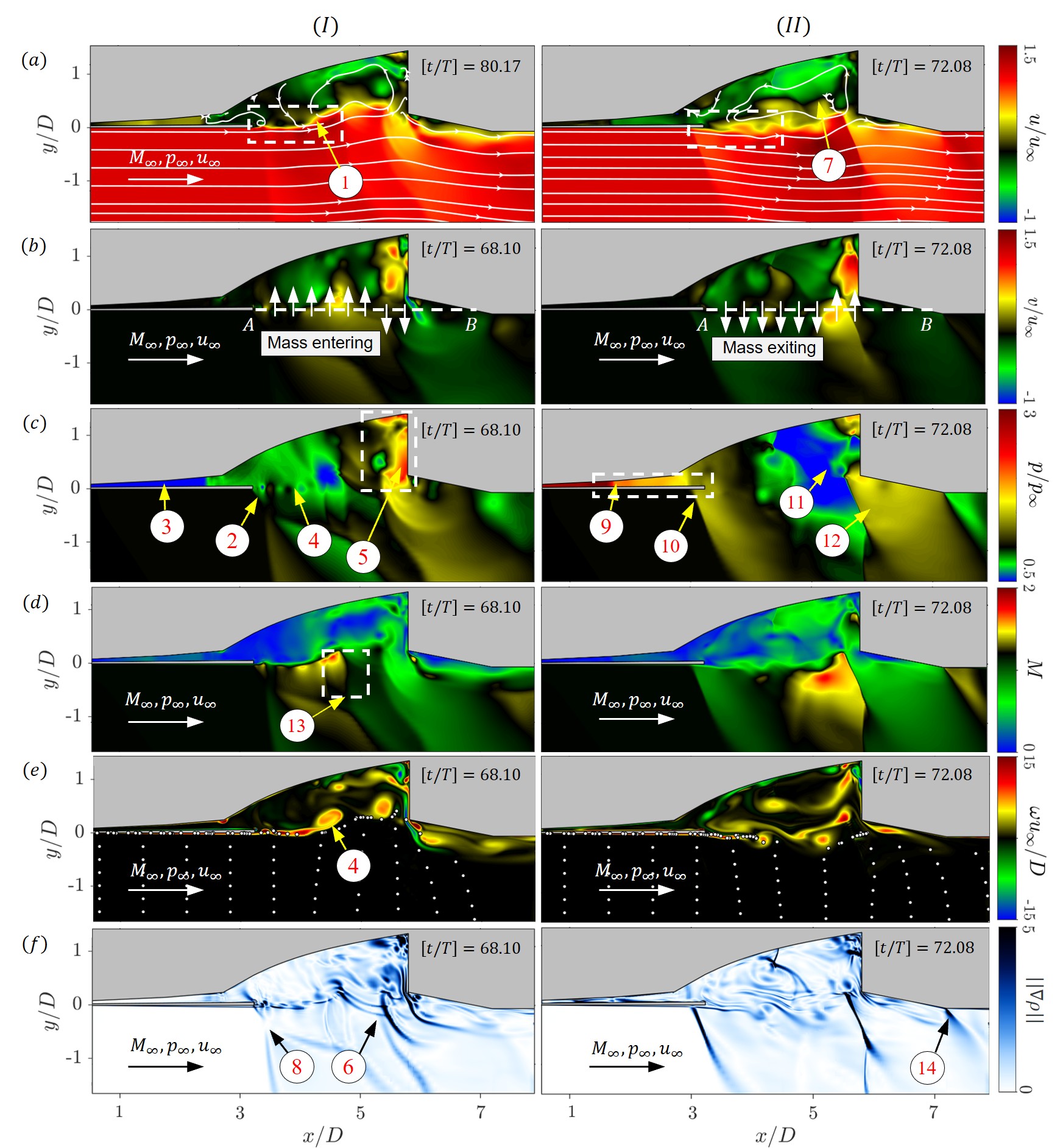}    \caption{\href{https://youtube.com/shorts/pF0axk4G8UI?feature=share}{(Multimedia View)} Instantaneous flow contours for the complex cavity-sub-cavity configuration at two different time stamps at $M_\infty=1.2$: (I) $[t/T] = 68.10$ (left) and (II) $[t/T] = 72.08$ (right), when the separated shear layer deflect inward and outward from the cavity opening (see line AB given by $[s_A/D,s_B/D] = [2.47,7.14]$) respectively. The normalized contours are shown for: (a) $x$-velocity ($u/u_\infty$), (b) $y$-velocity ($v/u_\infty$), (c) static pressure ($p/p_\infty$), (d) Mach number ($M$), (e) vorticity ($\omega D/u_\infty$), and (f) density gradient magnitude ($||\nabla \rho||$). Flow is from left to right. Key features observed during these stages include: (1) separated shear layer, (2) leading-edge expansion fan, (3) low-pressure region within sub-cavity, (4) vortical structures, (5) reflected compression wave,  (6) compression wave radiating into freestream, (7) recirculation bubble, (8) X-shaped reflected wave at leading edge, (9) high-pressure jet, (10) upstream-traveling compression wave at the leading edge, (11) low-pressure region in main cavity, (12) compression wave at trailing edge ramp, (13) shock wave, and (14) expansion fan downstream. The non-dimensionalizing variables used here are $D=3.8$ m, $u_\infty = 354.08$ m/$s$, and $p_\infty=$ 2.6kPa.}
    \label{fig:flow-physics}
\end{figure*}

 \begin{figure*}
    \centering
     \includegraphics[width=1\linewidth] {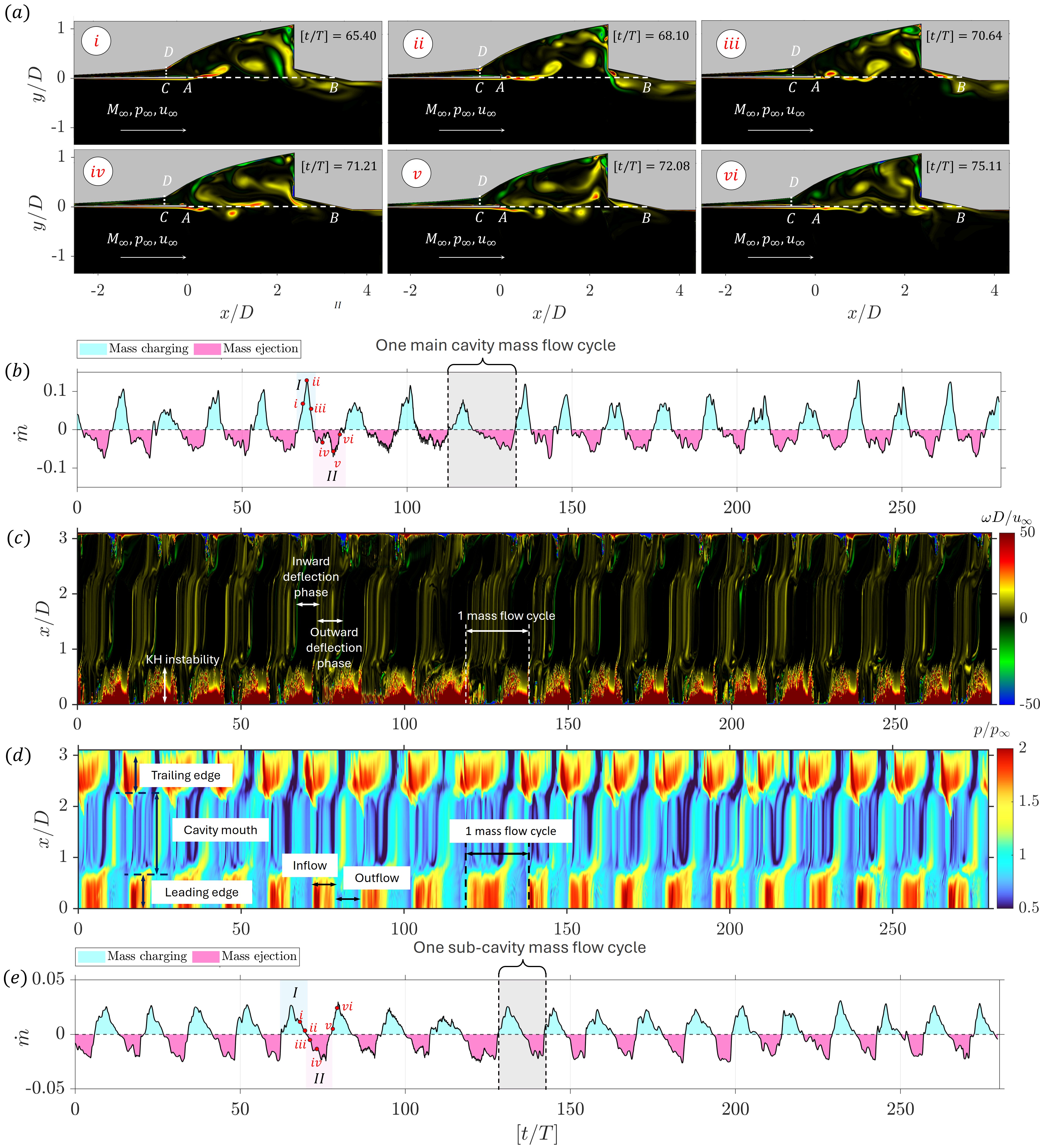}
    \caption{(a) Instantaneous snapshots of normalized vorticity contours illustrating the oscillatory motion of the shear layer as it deflects inward and outward from the cavity opening (see line AB given by $[s_A/D, s_B/D] = [2.47, 7.14]$) at $M_\infty=1.2$. Snapshots (i–iii) correspond to the inward deflection of the shear layer, while (iv–vi) represent the outward deflection; (b) Normalized mass flow rate ($\dot{m}$) across the primary cavity opening (line AB). The red markers indicate the corresponding flow instants (i-vi) shown in (a); (c-d) Evolution of non-dimensional vorticity and static pressure contours along the line AB over the considered non-dimensional flow time; (e) Normalized mass flow rate across the sub-cavity entrance ($s/D = 2.47$). The local mass flow rate is normalized using the freestream dynamic viscosity ($\mu_\infty$) and Reynolds number ($Re_{L,\infty}$), while the flow time is normalized with the freestream streamwise velocity $u_\infty =354.08$ m/$s$ and cavity depth $D=3.8$ m.}
   \label{fig:mass-flow}
\end{figure*}

\section{Results and Discussion} \label{sec: results}

\subsection{Data acquisition and post processing}
The spatial data for each time step is exported from ANSYS Fluent as an ASCII (American Standard Code for Information Interchange) text file. Cell connectivity information in the CGNS (Computational Grid Notation System) format is also exported at the end of the simulation. The scattered ASCII data, along with the connectivity information, are used to subsequently post-process spatial-temporal information using MATLAB\textsuperscript{\tiny\textregistered} scripts~\cite{ProcessingGuide2026}
for generating line plots, contour plots, and animation videos. 

\subsection{Accelerated time-stepping for self-sustained oscillation onset}

The onset of self-sustained oscillation in the cavity flow systems can be achieved using a staged time-stepping approach. Systematic time-step variations are employed during the simulation, and pressure fluctuations are monitored at probe locations. The shear-layer midpoint is found to be the most effective location for capturing the onset and evolution of these oscillations.

The adopted time-stepping scheme and the corresponding temporal evolution of pressure at the shear-layer midpoint are shown in Figure~\ref{fig:time-stepping}. The plot shows the flow development across five stages, labeled A–E, each represented by a different color corresponding to its simulation condition and duration. This approach drastically reduces the computational cost. The following events happen in the respective zones:

\begin{itemize}
\item Zone-A corresponds to the steady-state simulation in which a converged RANS solution is obtained using pseudo-time stepping. In Zone B, a velocity perturbation is introduced to disturb the steady RANS field and initiate the transition to large-scale oscillations. This perturbation is applied as a localized patch in the fluid domain to the $y$-velocity component with a magnitude of $0.1u_\infty$. The simulation is subsequently switched to a Detached Eddy Simulation (DES) module. 

\item Zone-C represents the growth phase of large-scale flow motion. During this stage, the flow evolves under coarse time stepping ($\Delta t/T = 4.8\times 10^{-3}$) and enters a self-sustained oscillatory state. Zone D corresponds to the stationary regime in which the self-sustained oscillations are fully established and persist without further qualitative changes in the flow behavior. 
 
\item In Zone-E, a finer time-step ($\Delta t/T = 1.2\times 10^{-3}$) is applied to accurately resolve the established oscillatory dynamics. The flow remains in the same stationary self-sustained oscillatory state during this phase. The non-dimensionalized time axis is referenced so that zero corresponds to the point at which the finer time step is applied, marking the beginning of Zone-E. All data used for post-processing and analysis are acquired when the simulation enters Zone-E.
\end{itemize}

Figure~\ref{fig:time-stepping}b presents the spectrogram of the pressure fluctuations recorded at the probe.  The spectrogram was computed with a window size of 100 samples and 80\% overlap. As the signal is acquired at varying time steps, the spectrogram is calculated by interpolating the signal with a uniform sampling rate of $f_s = 50$ kHz. The spectrogram is locally normalized at every time step to track the occurrence of the dominant tone. A strong discrete tone at $fD/u_\infty \approx 0.09$ is sustained throughout the time history. This indicates the presence of global unsteadiness in the complex cavity-sub-cavity system.

\subsection{Physics of the complex cavity-sub-cavity system}

In the previous section, the unsteady flow behavior was examined using pressure measurements at a single probe, $P_2$. This provided insight into the temporal characteristics of the unsteadiness at a specific location. However, such point-based information alone cannot capture how this unsteadiness develops and propagates throughout the computational domain. It is essential to examine the spatial distribution of the flow field to understand how local unsteadiness manifests and evolves across the entire domain. The global unsteadiness is therefore investigated using instantaneous flow contours of key variables such as $x-$velocity ($u$), $y-$velocity ($v$), static pressure ($p$), Mach number ($M_\infty$), vorticity ($\omega$), and density gradient magnitude ($||\nabla \rho||$) as shown in Figure~\ref{fig:flow-physics}. These flow fields collectively illustrate the flow evolution and interaction of coherent structures from the separated shear layer that govern the unsteady behavior of the complex cavity-sub-cavity system (see the associated \href{https://youtu.be/eR1i7_N9yTc}{Multimedia View} to visualize the global unsteadiness). 

When the freestream flow encounters the complex cavity-sub-cavity system, several characteristic flow features emerge at transonic conditions ($0.8\leq M_\infty\leq 1.2$), as shown in Figure~\ref{fig:flow-physics}. These include the formation of a separated shear layer (feature 1) that spans the cavity opening from the leading edge to the trailing edge, the generation of compression and expansion waves, the emission of acoustic waves, and the establishment of a recirculation region within the cavity. The interaction between these flow features creates a feedback loop that sustains the global unsteadiness within the cavity-sub-cavity system. Consequently, the shear layer undergoes periodic oscillations, deflecting inward and outward across the cavity opening (line AB, $[x_A/D,y_A/D] = [0,0] $ and $[x_B/D,y_B/D] = [3.8,0]$ in Figure~\ref{fig:flow-physics}b). This regulates mass ingestion/ejection within the cavity-sub-cavity system and modulates the unsteadiness throughout the domain. 

Temporal mass-flow rate across the cavity opening (line AB) is evaluated to quantify this behavior. In addition, the streamwise mass-flow rate entering the sub-cavity is computed along a vertical line CD bounded between the walls at $[x_c/D,y_c/D]=[-0.0.3,0.053]$ and $[x_D/D,y_D/D]=[-0.0.3,0.083]$ (see Figure~\ref{fig:mass-flow}e) to understand the mass flow coupling in the cavity-sub-cavity system. The equation for calculating the non-dimensionalized mass flow rate is given as,
\begin{equation}
\dot{m}(t) = \frac{1}{\mu_\infty Re_{L,\infty}} \sum_{j=1}^{N} \rho_j \, U_j \, \Delta x_j,
\end{equation}
where $\rho_j$ and $U_j$ denote the local density and velocity (streamwise or transverse velocity based on the measurement location) at the $j^{th}$ point along the measurement line, respectively. $\Delta x_j$ represents the local grid spacing between adjacent points along the same line. 

Figure~\ref{fig:mass-flow}a shows a sequence of instantaneous vorticity snapshots (i-vi) of one complete mass inflow-outflow cycle (see Figure~\ref{fig:mass-flow}b), during which the shear layer undergoes periodic inward and outward deflections relative to the cavity opening (line AB). Figure~\ref{fig:mass-flow}b shows the temporal variation of the mass-flow rate across the line AB. The shaded region highlights one representative mass inflow/outflow phase. Positive and negative mass flux clearly identify alternating phases of mass ingestion into the cavity and mass ejection from it, respectively. Figures~\ref{fig:mass-flow}c-d show the corresponding time-resolved vorticity and pressure contours, respectively, evaluated along the same line AB during the mass-flow cycle. 

During the inflow phase, when the shear layer deflects inward along line AB (Figure~\ref{fig:mass-flow}a, i–iii), the time-resolved vorticity contours (Figure~\ref{fig:mass-flow}c) reveal intense vorticity concentrations near the cavity leading edge. This elevated vorticity is indicative of the cavity developing KH-type instability as the separated shear rolls up into the cavity, breaking down into small- and large-scale vortical structures, and then convecting downstream. Simultaneously, the pressure contours shown in Figure~\ref{fig:mass-flow}d indicate the formation of a pronounced low-pressure region at the leading edge, accompanied by a sharp pressure gradient extending downstream along the cavity length. This pressure signature suggests the presence of a compression wave propagating along the cavity during the inflow phase. In addition, the trailing edge experiences a transient region of elevated pressure during this phase.

As the mass-flow cycle transitions to the outflow phase, the shear layer deflects outward from line AB (Figure~\ref{fig:mass-flow}a, iv–vi). Correspondingly, the vorticity magnitude near the cavity leading edge decreases, as observed in Figure~\ref{fig:mass-flow}c. The pressure contours in Figure~\ref{fig:mass-flow}d show a region of elevated pressure near the cavity leading edge, indicating that mass exiting the cavity opening does so at relatively high pressure at this location. Consequently, the cavity leading edge experiences alternating low- and high-pressure states over successive inflow and outflow phases. These periodic pressure fluctuations are directly linked to the unsteady oscillation of the separated shear layer and govern the unsteadiness observed across the cavity-sub-cavity system. In addition to the primary cavity, the sub-cavity region also exhibits periodic mass inflow and outflow, as evidenced by the mass flow rate measured across line CD and shown in Fig.~\ref{fig:mass-flow}e. This behavior highlights the strong coupling between the primary cavity and the sub-cavity flow cycles. 

Two representative instances, marked as I and II in Fig.~\ref{fig:flow-physics}, corresponding to non-dimensional times $[t/T] = 68.10$ and $[t/T] = 72.08$, respectively, are selected to correlate the instantaneous flow structures with the inflow and outflow phases. During the shear layer's inward-deflection (Figure \ref{fig:flow-physics}I a–d), the freestream encounters a geometric expansion at the leading edge, forming an expansion fan (feature 2). Within this stage, the sub-cavity experiences a pronounced low-pressure zone (feature 3). The separated shear layer breaks down into small and large-scale vortical structures (feature 4) that convect downstream and impinge on the trailing edge. Their impingement generates a strong compression wave (feature 5), which produces a localized pressure rise near the cavity neck. This compression wave radiates into the freestream (feature 6) and perturbs the shear-layer. Concurrently, a weaker, upstream-traveling acoustic pulse propagates through the recirculating flow (feature 7) inside the cavity. Portions of this pulse reflect off the curved bottom wall, and subsequent interactions between the reflected waves and the leading-edge expansion fan form an “X-shaped” reflected wave (feature 8) pattern near the cavity’s leading edge. These upstream-moving disturbances further perturb the separated shear layer, initiating the next roll-down cycle. 

The subsequent outward deflection of the shear layer (Figure \ref{fig:flow-physics}II a–d) across the cavity opening is shown in Figure \ref{fig:mass-flow}a. This deflection of the shear layer coincides with the ejection of a high-pressure jet (feature 9) from the sub-cavity, produced by fluid that entered during the previous inflow cycle and is reflected from the sub-cavity wall. The jet perturbs the shear layer near the leading edge, generating an oblique shock (feature 10) as it encounters the incoming freestream. Additionally, the presence of the inclined trailing-edge ramp causes the impinging shear layer to separate and reattach, producing a secondary compression wave (feature 12) throughout. The cavity is observed trapping a low-pressure pocket between two high-pressure regions (features 10 and 12). This periodic phenomenon observed within the cavity establishes a self-sustaining feedback loop, generating pressure fluctuations in the cavity-sub-cavity system. 

The observed key flow feedback across the complex cavity-sub-cavity system closely resembles the acoustic feedback loop proposed by Rossiter~\cite{rossiter1964}. According to the modified Rossiter’s model~\cite{heller1971}, the dominant oscillation frequencies in cavity flows can be predicted using an empirical relation between the non-dimensional frequency (Strouhal number, $St$), the convective motion of shear-layer vortices, and the acoustic feedback within the cavity. The empirical relation is expressed as,
\begin{equation}
St_L = \frac{fL}{u_\infty} = \frac{n - \alpha}{\dfrac{1}{\kappa} + \dfrac{M_\infty}{\left({1 + \frac{\gamma - 1}{2}M_\infty^2}\right)^{1/2}}},
\label{eq:rossiter eqn}
\end{equation}
where $n$ is the mode number, $\alpha = 0.25$ is the phase delay associated with the feedback loop, $\kappa =0.57$ is the ratio of the convective velocity of the vortical structures to the freestream velocity, $M_\infty$ is the freestream Mach number, $L$ is the cavity length, and $\gamma$ is the ratio of specific heat.

\begin{figure}
    \centering
   \includegraphics[width=1\linewidth] {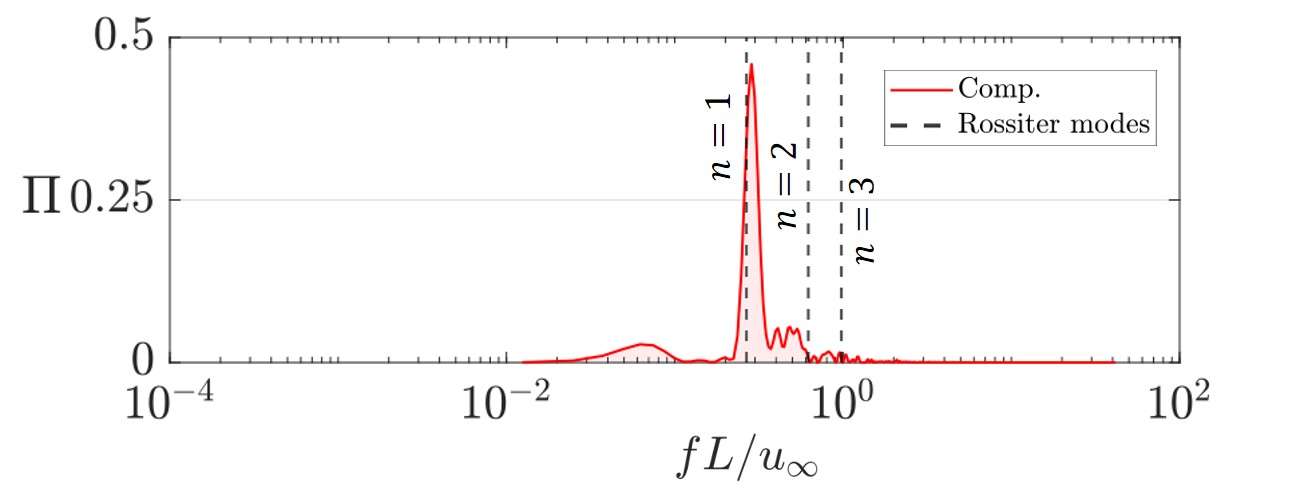}
    \caption{Non-dimensional Power Spectral Density ($\Pi = fG_{xx}/p_r^2$) versus non-dimensional frequency ($fD/u_\infty$) for the static pressure signal at probe $P_2$ (red) in a flow of $M_\infty=1.2$. Black dashed lines indicate the first three Rossiter-mode frequencies ($n$ = 1, 2, and 3) predicted by Eq.~\ref{eq:rossiter eqn}. The non-dimensionalizing variables used here are $L=11.78$ m, $u_\infty = 354.08$ m/$s$, and $p_r=$ 1kPa.}
    \label{fig:rossiter}
\end{figure}

Figure~\ref{fig:rossiter} shows the spectrum obtained from FFT analysis of the static pressure measured at the probe $P_2$ (see Figure~\ref{fig:computational_domain}). The black vertical line in the spectrum denotes the frequency predicted by Rossiter's~\cite{rossiter1964} equation for the corresponding flow conditions. The predicted first Rossiter mode frequency closely coincides with the dominant frequency obtained from the numerical simulation. This agreement implies that a complex cavity-sub-cavity system is governed by acoustic-hydrodynamic feedback between the shear layer and the cavity geometry.

\begin{figure*}
    \centering
    \includegraphics[width=1\linewidth] {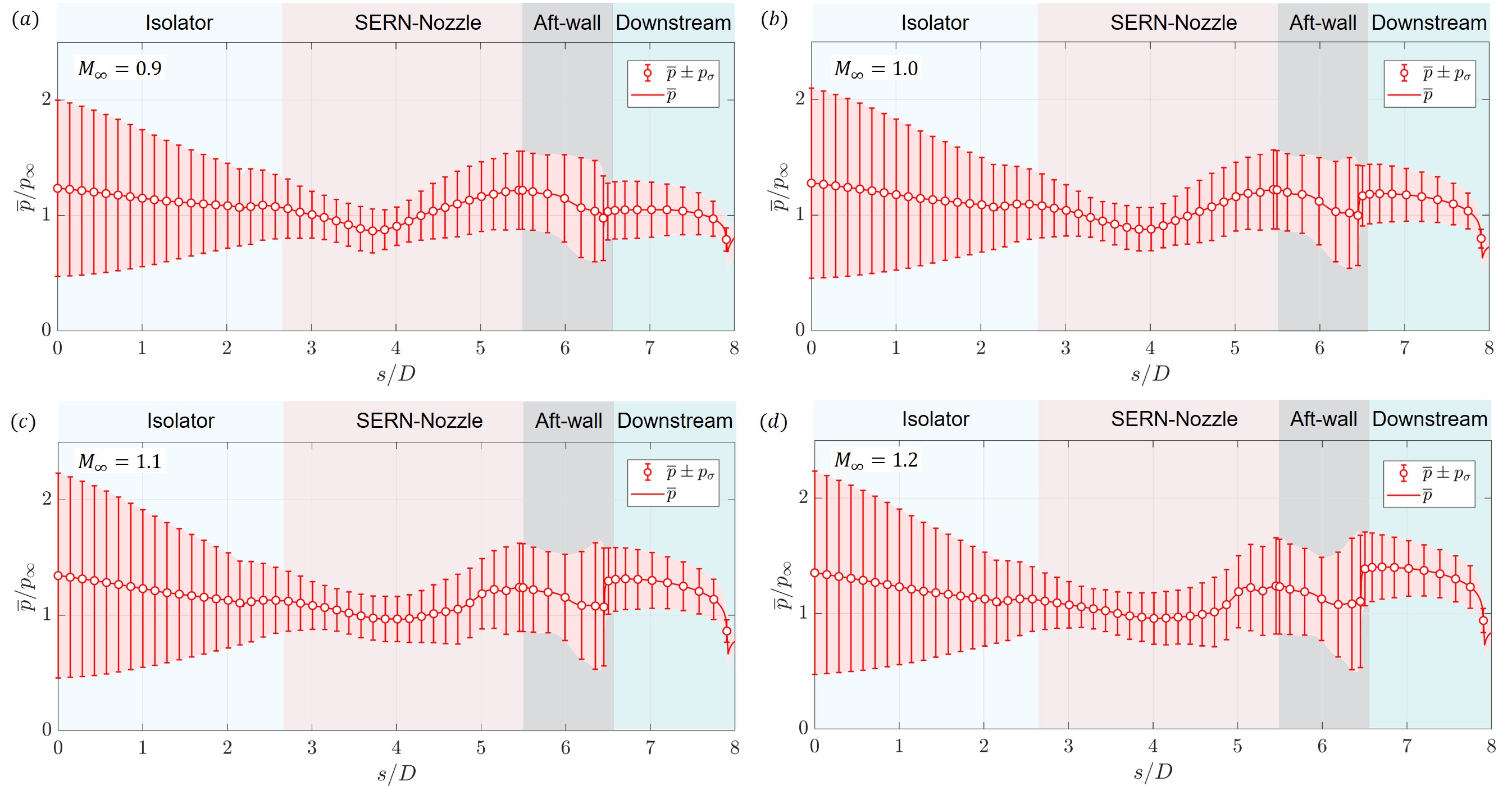}
    \caption{Plots of non-dimensionalized time-averaged wall-static pressure ($\overline{p}/p_\infty$ as markers) and standard deviation of the wall-static pressure fluctuations ($p_\sigma /p_\infty$ as error bars) over the different parts of the scramjet surface for different freestream Mach numbers: (a) $M_\infty = 0.9$, (b) $M_\infty = 1.0$, (c) $M_\infty =1.1$, and (d) $M_\infty = 1.2$. The plot is done on the surface coordinate system $[s/D]$ as mentioned in Figure \ref{fig:computational_domain}. The respective scramjet surfaces as shown in Figure \ref{fig:problem_statement2}e are indicated in different background colors: isolator (light blue), SERN nozzle (light red), aft-wall (light gray), and downstream zone (bright blue). The non-dimensionalizing variables used here are $D=3.8$ m and $p_\infty=$ 2.6kPa.}
    \label{fig:rms_plot}
\end{figure*}

\begin{figure*}
    \centering
    \includegraphics[width=1\linewidth] {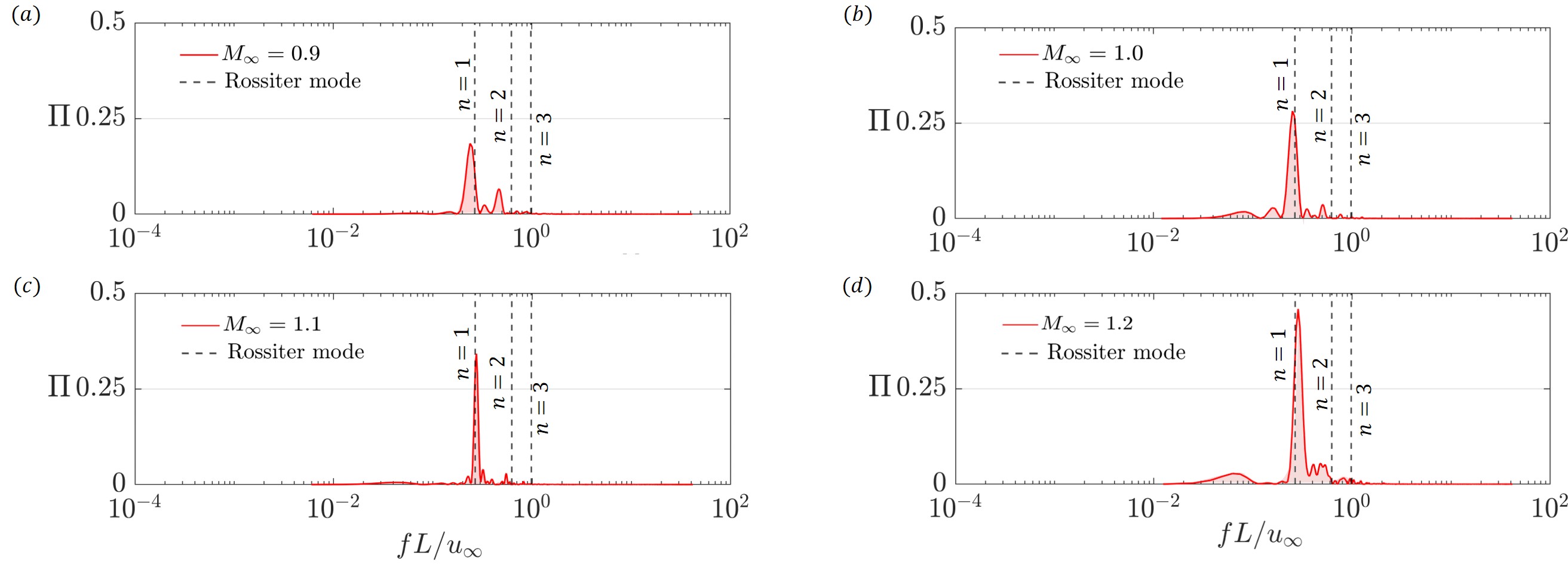}
    \caption{Non-dimensional power spectral density ($\Pi=fG_{xx}(f)/p_r^2$) of static pressure fluctuations at $P_1$ (sub-cavity end-wall) of the actual cavity-sub-cavity geometry for different freestream Mach numbers: (a) $M_\infty = 0.9$, (b) $M_\infty=1.0$, (c) $M_\infty=1.1$, and (d) $M_\infty=1.2$. The spectra are shown as a function of the non-dimensional frequency, $fL/u_\infty$. Dashed vertical lines indicate the first three Rossiter modes ($n$=1, 2, and 3).  The non-dimensionalizing variables used here are $L=11.78$ m, $u_\infty = 354.08$ m/$s$, and $p_r=$ 1kPa.}
    \label{fig:mach_fft_plot}
\end{figure*}

\subsection{Flow behavior across transonic Mach numbers}
\begin{table*}
\centering
\caption{Static pressure statistics at $P_1$ (sub-cavity end-wall) location of the actual cavity-sub-cavity geometry for different freestream Mach numbers (Note: mean pressure - $\overline{p}/p_\infty$, standard deviation of pressure - $p_{\sigma}/p_\infty$, non-dimensional frequency - $fL/u_\infty$, non-dimensional spectral power - $\Pi=fG_{xx}(f)/p_r^2$, and `\% change' refer to variation relative to $M_\infty= 0.9$). The non-dimensionalizing variables used here are $L=11.78$ m, $u_\infty = 354.08$ m/$s$, $p_\infty =$2.6kPa, and $p_r=$ 1kPa.}
\begin{ruledtabular}
\begin{tabular}{lcccccccc}
$M_\infty$ & $\overline{p}/p_\infty$  & $p_{\sigma}/p_\infty$ & $fL/u_\infty$ & \% change & $\Pi$ & \% change \\ \midrule
0.9 & 1.45  & 0.18 & 0.1 & --   & 0.14 & -- \\
1.0 & 1.50  & 0.18 & 0.1 & 0.4  & 0.30 & 15.5 \\
1.1 & 1.60  & 0.20 & 0.1 & 0.4 & 0.36 & 22.0 \\
1.2 & 1.62  & 0.20 & 0.1 & 0.3 & 0.45 & 31.5 \\
\end{tabular}
\end{ruledtabular}
\label{tab:mach_pressure}
\end{table*}
The pressure-loading characteristics of the complex-cavity–sub-cavity system are examined to determine their dependence on the freestream Mach number. Figure~\ref{fig:rms_plot} shows the non-dimensional time-averaged static pressure distribution ($\overline{p}/p_\infty$) along the cavity-sub-cavity surface for $M_\infty = 0.9$–$1.2$. At the sub-cavity end wall ($s/D = 0$, see Figure \ref{fig:computational_domain}) as the freestream  Mach number increases from subsonic to supersonic, the mean pressure loading increases by 15-30\%. For $M_\infty = 1.1$ and $M_\infty = 1.2$, the pressure loading experienced on the end-wall is $\approx1.6$ times higher than the freestream pressure ($p_\infty$). Along the sub-cavity length ($l$), the mean pressure decreases gradually due to changes in cross-sectional area. This is followed by a dip in pressure near the cavity's leading edge ($s/D = 2.71$) attributed to mass entrainment/ejection through the cavity's leading edge as a result of shear layer oscillation. From the cavity leading edge to the mid-cavity region ($s/D = 3.4$), a sustained low-pressure plateau is observed in the pressure distribution, indicating the establishment of a persistent low-pressure environment over time. This plateau is most pronounced for $M_\infty = 0.9$, suggesting that the subsonic flow regime experiences substantially lower mean pressure levels in the mid-cavity region compared to the higher Mach number cases. Beyond this region, the pressure gradually increases along the remaining length of the cavity. The aft wall experiences elevated pressure loading for all Mach numbers. For $M_\infty = 0.9$ and $M_\infty = 1.0$, the pressure rise is relatively steady, whereas for $M_\infty = 1.1$ and $M_\infty = 1.2$, the aft wall encounters increased pressure fluctuations. These fluctuations are primarily attributed to interactions between the ramp-induced shock system and the unsteady shear layer, which become more pronounced at supersonic Mach numbers. Further downstream, in the ramp region ($5.4 \leq s/D \leq 6.5$), flow separation and subsequent reattachment at the trailing edge promote pressure recovery of the separated shear layer flow. The presence of an oblique shock at the ramp trailing edge (feature 11, see Figure~\ref{fig:flow-physics}II-c) further enhances the local pressure rise across this region, thereby aiding pressure recovery. The magnitude of pressure recovery increases with Mach number, with $M_\infty = 1.2$ exhibiting the highest recovery of approximately 21\%. This trend reflects the role of stronger shock structures at higher Mach numbers in enhancing pressure recovery within the cavity–sub-cavity system. In contrast, the subsonic case ($M_\infty = 0.9$) exhibits weaker pressure recovery, consistent with the absence of dominant shock structures.

Figure~\ref{fig:mach_fft_plot} presents the spectrum obtained from FFT analysis of the static pressure signal measured at the sub-cavity end wall ($s/D = 0$). Table~\ref{tab:mach_pressure} summarizes the variation of non-dimensional pressure characteristics at the sub-cavity end wall with increasing freestream Mach number. The peak pressure power ($\Pi$) increases monotonically with Mach number, with the peak pressure level rising by approximately 31.5\% for $M_\infty = 1.2$, indicating significantly higher pressure loading on the sub-cavity end wall. Whereas, the dominant non-dimensional frequency ($fL/u_\infty$) exhibits only a weak dependence on Mach number, showing a marginal increase of about 0.3–0.4\% between $M_\infty = 0.9$ and $M_\infty = 1.2$. Such small variations are negligible and therefore do not raise concerns regarding numerical resolution or temporal accuracy.

\begin{figure}
    \centering
    \includegraphics[width=1\linewidth] {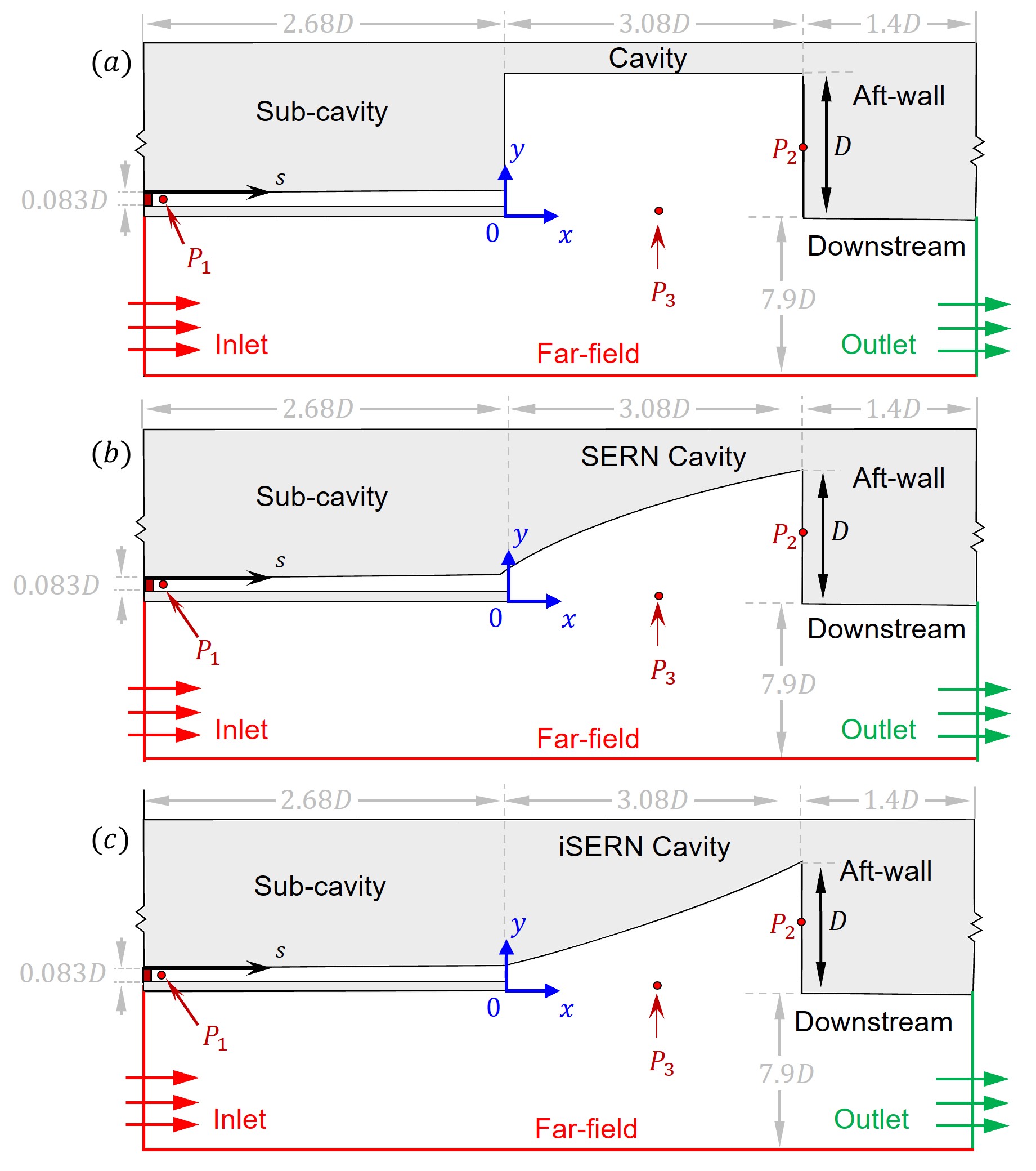}
    \caption{Computational domain for (a) Baseline geometry (BG) with a rectangular-shaped cavity, (b) Single Expansion Ramp Nozzle (SERN)-shaped cavity (SG), and (c) Inverted SERN-shaped cavity (IG). All geometries are parameterized by cavity depth ($D$) and are subjected to a freestream flow with $M_\infty=1.2$. The probe locations $P_1$, $P_2$, and $P_3$ are also shown.}
    \label{fig:schematic_shape}
\end{figure}

\begin{figure*}
    \centering
    \includegraphics[width=1\linewidth] {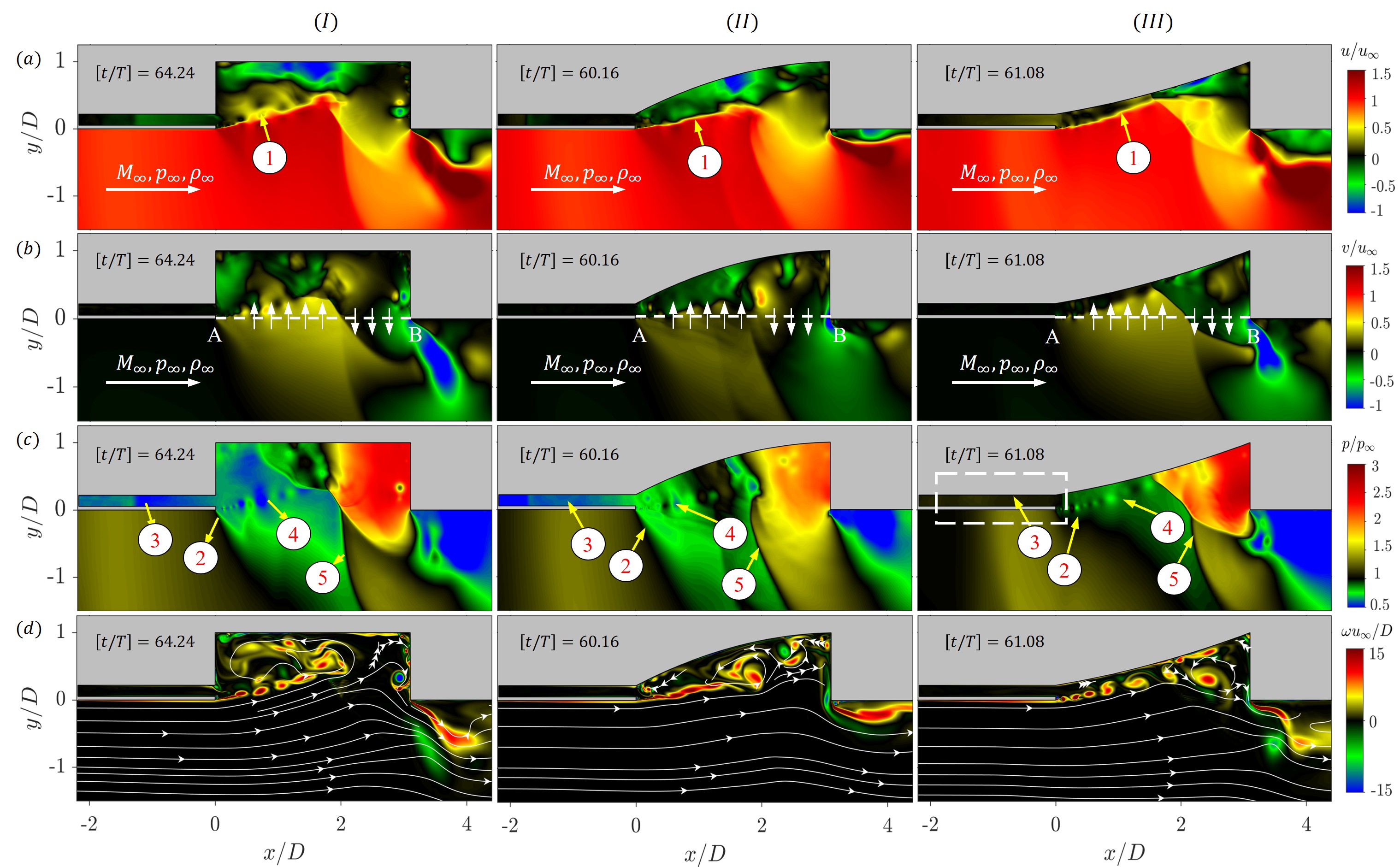}
    \caption{\href{https://youtu.be/jZW4gtCghVY}{(Multimedia View)} Instantaneous contour plots at $M_\infty=1.2$ for (I) Baseline (BG), (II) SERN (SG), and (III) iSERN (IG) geometry shown in Figure~ \ref{fig:schematic_shape} at the time instant when the separated shear layer deflects upward. Normalized contours are shown for (a) $x$-velocity ($u/u_\infty$), (b) $y$-velocity ($v/u_\infty$), (c) static pressure ($p/p_\infty$), and (d) vorticity ($\omega/u_\infty D$). Flow is from left to right. Key features include: (1) separated shear layer, (2) leading-edge expansion fan, (3) low-pressure region in sub-cavity, (4) vortical structures, and (5) reflected compression wave from aft-wall. The non-dimensionalizing variables used here are $D=3.8$ m, $u_\infty = 354.08$ m/$s$, and $p_\infty=$ 2.6kPa.}
    \label{fig:shear_upward}
\end{figure*}

\begin{figure*}
    \centering
    \includegraphics[width=1\linewidth] {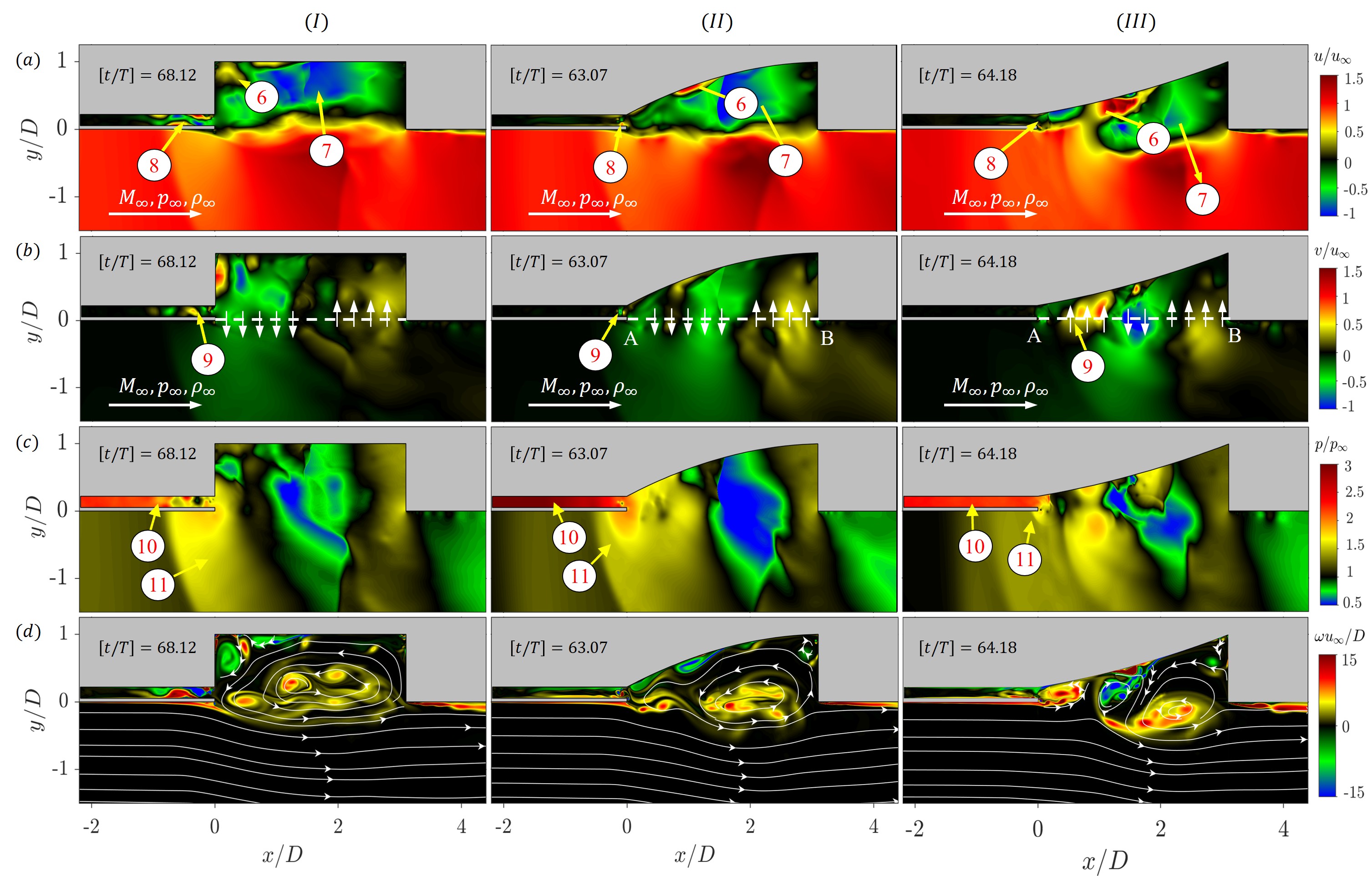}
    \caption{\href{https://youtu.be/xO3K23wDFiA}{(Multimedia View)} Instantaneous contour plots at $M_\infty=1.2$ for (I) Baseline (BG), (II) SERN (SG), and (III) iSERN (IG) geometry shown in Figure~ \ref{fig:schematic_shape} at the time instant when the separated shear layer deflects downward. Normalized contours are shown for (a) $x$-velocity ($u/u_\infty$), (b) $y$-velocity ($v/u_\infty$), (c) static pressure ($p/p_\infty$), and (d) vorticity ($\omega/u_\infty D$). Flow is from left to right. Key features include: (6) corner vortices, (7) large recirculation bubble, (8) sonic flow at the sub-cavity neck, (9) vertical flow at the sub-cavity neck, (10) high-pressure flow exiting the sub-cavity region, and (11) upstream-traveling compression wave from the leading edge of the cavity. The non-dimensionalizing variables used here are $D=3.8$ m, $u_\infty = 354.08$ m/$s$, and $p_\infty=$ 2.6kPa.}
    \label{fig:shear_outward}
\end{figure*}

\begin{figure*}
    \centering
    \includegraphics[width=1\linewidth] {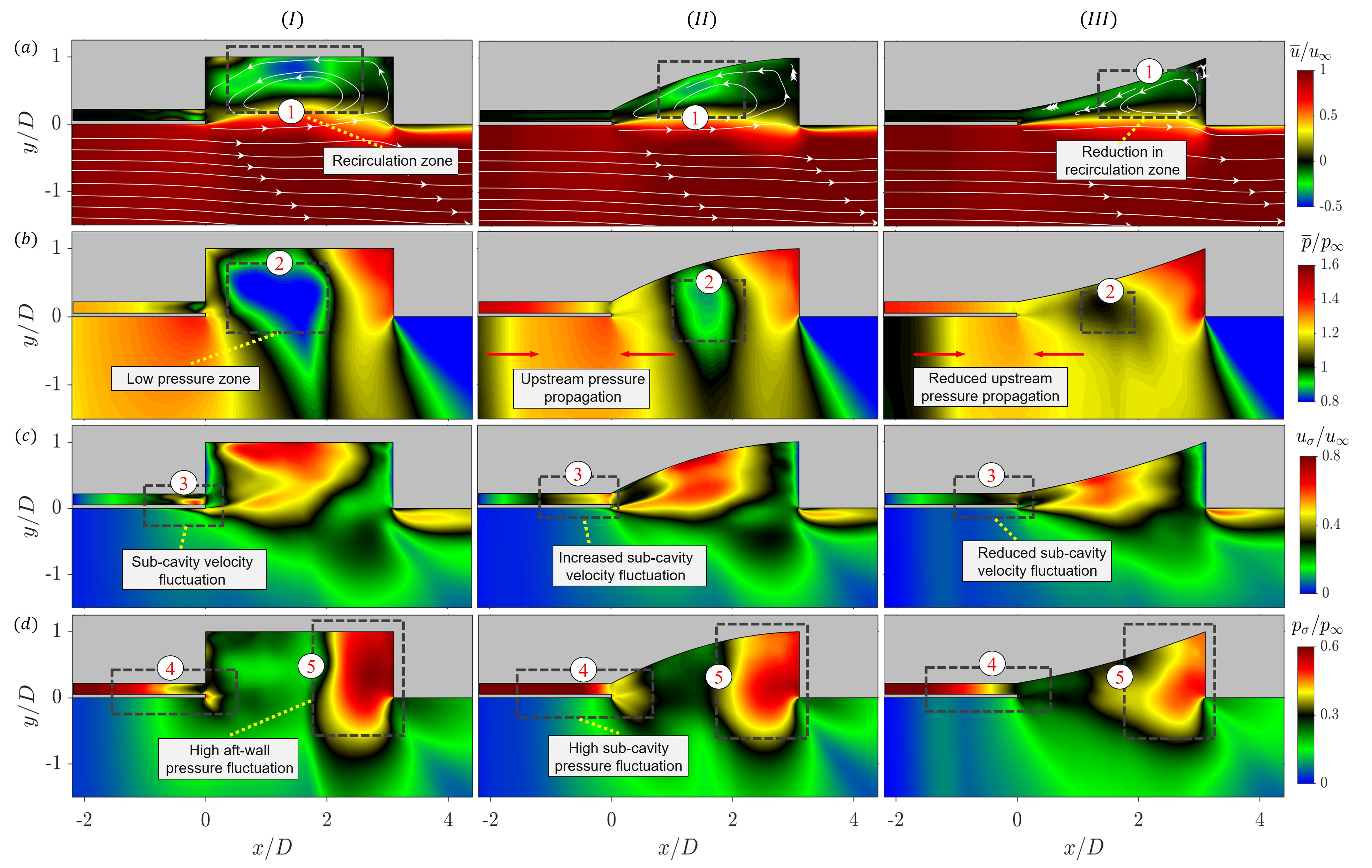}
    \caption{Non-dimensionalized contour plots at $M_\infty=1.2$ showing (a) time-averaged velocity, (b) time-averaged static pressure, (c) standard deviation of velocity, and (d) standard deviation of pressure for three different geometries: (I) Baseline (BG), (II) SERN (SG), and (III) iSERN (IG). Key flow features: (1) recirculation zone, (2) low-pressure zone, (3) sub-cavity velocity fluctuations, (4) sub-cavity high-pressure fluctuations, and (5) aft-wall high-pressure fluctuations. The non-dimensionalizing variables used here are $D=3.8$ m, $u_\infty = 354.08$ m/$s$, and $p_\infty=$ 2.6kPa.}
\label{fig:cavity_shape_shear_mean}
\end{figure*}

\begin{figure*}
    \centering
    \includegraphics[width=1\linewidth] {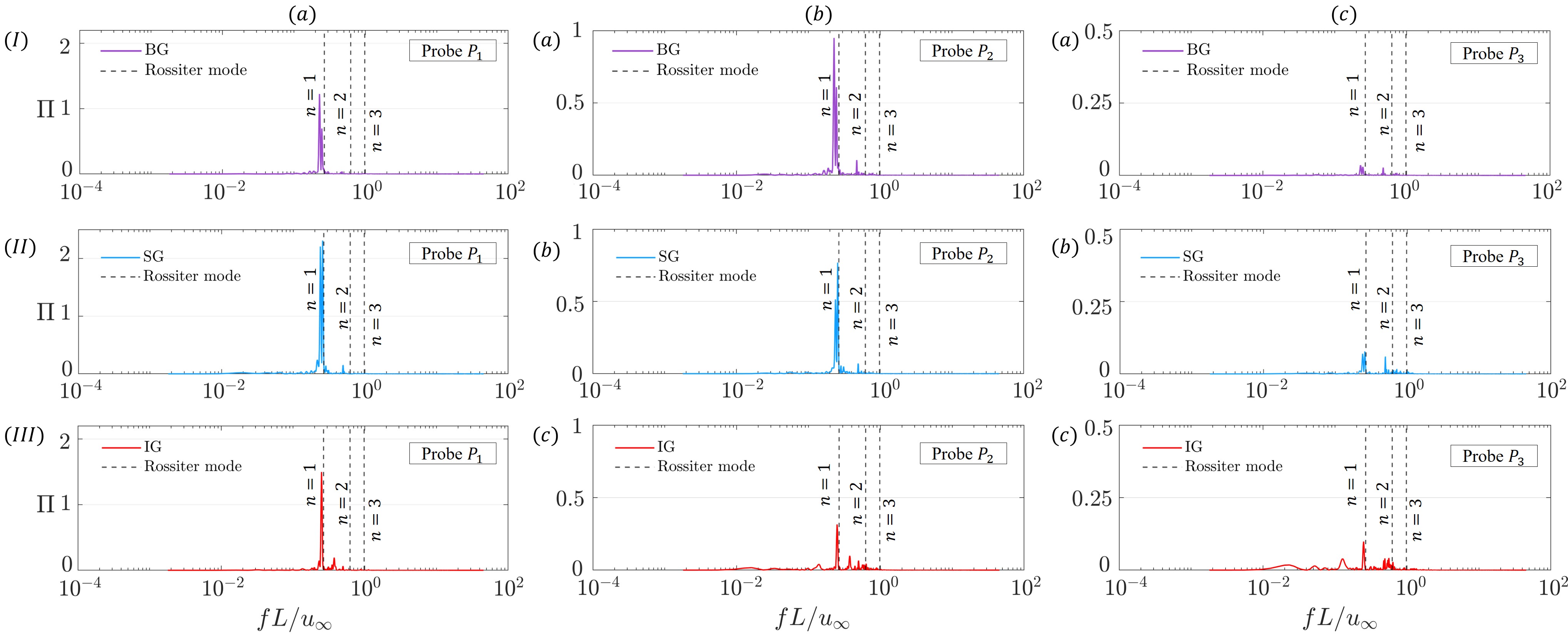}
    \caption{Non-dimensional power spectral density ($\Pi = fG_{xx}/p_r^2$) of static pressure fluctuations for the geometries shown in Figure~\ref{fig:schematic_shape}at probe locations: (a) $P_1$ (sub-cavity end wall), (b) $P_2$ (aft-wall), and (c) $P_3$ (shear-layer midpoint, for the (I) BG (top row),  (II) SG (middle row), and (III) IG (bottom row). The spectra are shown as a function of the non-dimensional frequency, $fL/u_\infty$. Dashed vertical lines indicate the first three Rossiter modes ($n$ = 1, 2, and 3). The non-dimensionalizing variables used here are $L=11.78$ m, $u_\infty = 354.08$ m/$s$, and $p_r=$ 1kPa.}
    \label{fig:cavity_shape_fft2}
\end{figure*}

\subsection{Cavity topology}
The internal geometry plays a significant role in the unsteady dynamics observed in cavity flows~\cite{gharib1987}. This section highlights the effects of varying the topology of the cavity-sub-cavity system. Three geometry topologies were used for this purpose. A baseline geometry (BG), SERN (Single Expansion Ramp Nozzle) geometry (SG), and inverted-SERN geometry (IG) were used for the study. The SERN~\cite{capone1992,re1982} topology was selected because it closely resembles the two-dimensional or quasi-two-dimensional nozzle geometries and the inverted SERN geometry represents the characteristic configuration encountered in aerospike nozzle~\cite{aerospike_review2024,design_aerospike2020} systems. The detailed schematics of the geometries used are provided in Figure~\ref{fig:schematic_shape}.

Each geometry incorporates a deep sub-cavity with the same $l/d$ ratio, isolating the main cavity topology as the primary variable. This allows a direct assessment of how cavity topology influences the unsteady flow behavior. The underlying flow physics observed for all geometries is consistent with the mechanism discussed in Section~\ref{sec: results}b. Figure~\ref{fig:shear_upward} and Figure~\ref{fig:shear_outward} shows the instantaneous field contours during inward and outward deflections of the shear layer, respectively. Due to differences in topology, the available volume for mass entrainment varies among the three geometries. The available volume is largest for BG and smallest for IG.

\begin{table}
\centering
\caption{Mass flux statistics for geometries shown in Figure~\ref{fig:schematic_shape} across the cavity opening (see line AB given by $[s_A/D, s_B/D] = [2.35, 7.02]$ in Figure~\ref{fig:shear_upward}). (Note: mean mass-flux - $\overline{\dot{m}}$, standard deviation of mass flux - $\dot{m}_{\sigma}$, ratio between mass entering ($\dot{m_i}$) to mass leaving ($\dot{m_e}$) the cavity opening - $\eta$, skewness factor - $\beta_1$ and kurtosis factor - $\beta_2$.}
\begin{ruledtabular}
\begin{tabular}{lccccc}
Geometry 
& $\overline{\dot{m}}$ 
& $\dot{m}_{\sigma}$ 
& $\eta_m$ 
& $\beta_1$ 
& $\beta_2$ \\ \midrule

Baseline (BG) 
& 0.47 & 6.93 & 1.5 & 0.7 & 2.7 \\

SERN (SG)     
& 0.38 & 6.90 & 1.2 & $-0.1$ & 2.8 \\

iSERN (IG)    
& 0.32 & 5.41 & 1.1 & 0.3 & 3.1 \\

\end{tabular}
\end{ruledtabular}
\label{tab:mass_stats}
\end{table}

\begin{figure*}
    \centering
    \includegraphics[width=1\linewidth] {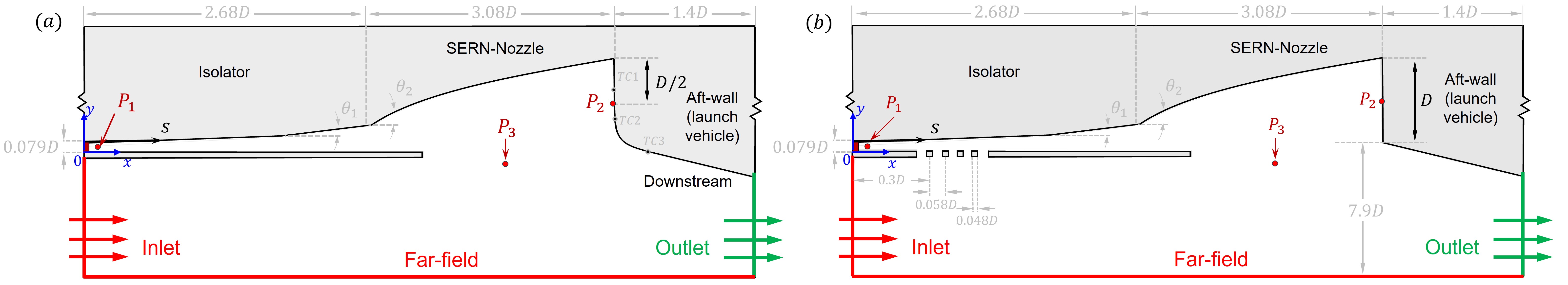}
    \caption{Schematic showing the computational domain extent and the geometry details for two different control cases: (a) case-1 (C1) with a chamfered aft-wall and (b) case-2 (C2) with a slotted sub-cavity wall for a flow at $M_\infty = 1.2$. The geometries are parameterized using cavity depth $D$.}
    \label{fig:cavity_passive_control}
\end{figure*}

\begin{figure*}
   \centering
    \includegraphics[width=1\linewidth] {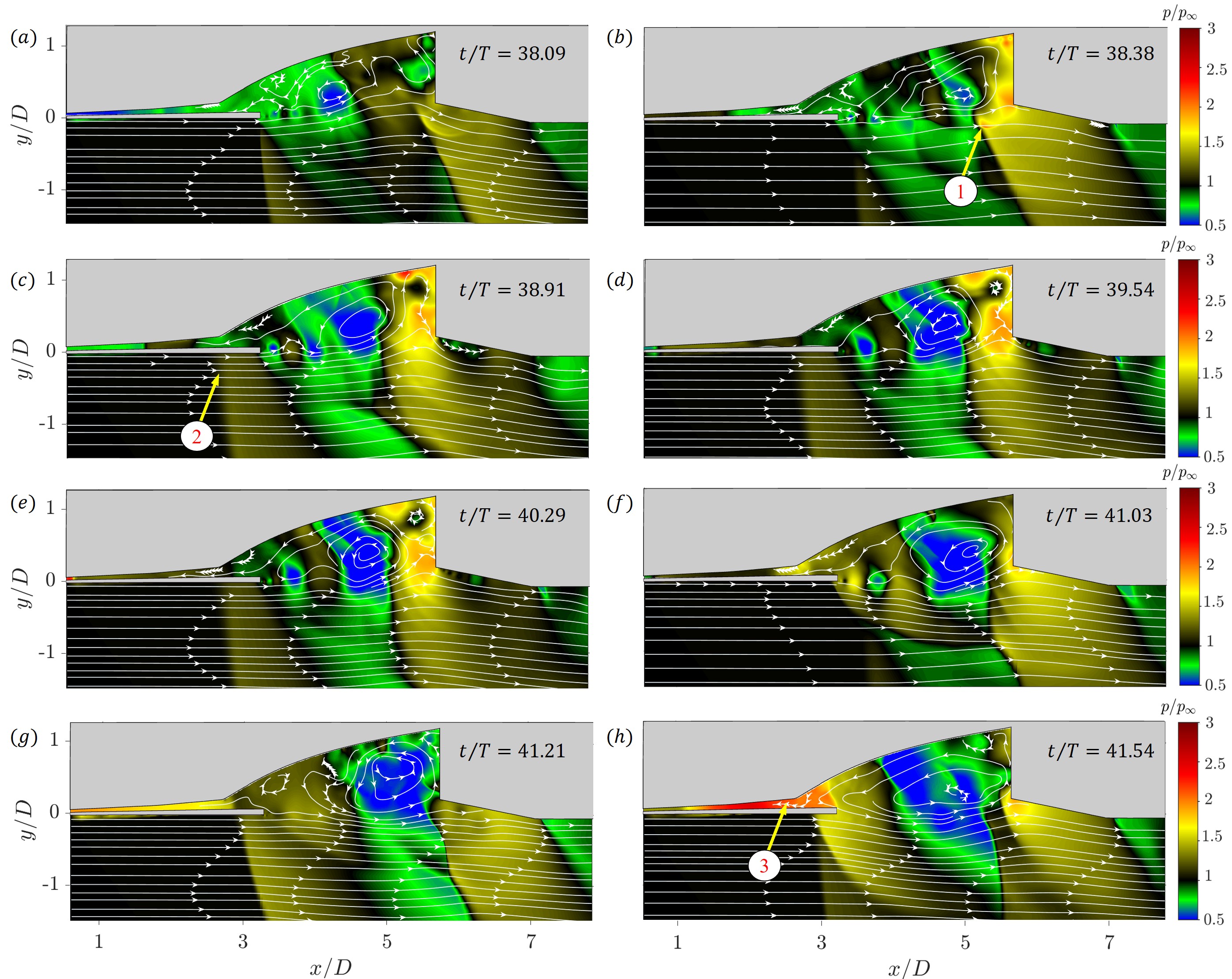}
    \caption{Instantaneous pressure contour plots at $M_\infty=1.2$ of the complex cavity-sub-cavity system (C0) over one Rossiter oscillation cycle highlight key flow features: (1) High-pressure intensity concentrated at the aft wall, (2) Upstream propagating compression wave from the leading edge of the cavity and (3) Ejection of high-pressure flow from the sub-cavity. The superimposed streamlines further illustrate the dynamic exchange across the cavity opening, capturing both inflow and outflow processes, along with the formation of large recirculation zone within the cavity.  The non-dimensionalizing variables used here are $D=3.8$ m and $p_\infty=$ 2.6kPa.}
    \label{fig:passive_control_contour1}
\end{figure*}

\begin{figure*}
   \centering
   \includegraphics[width=1\linewidth] {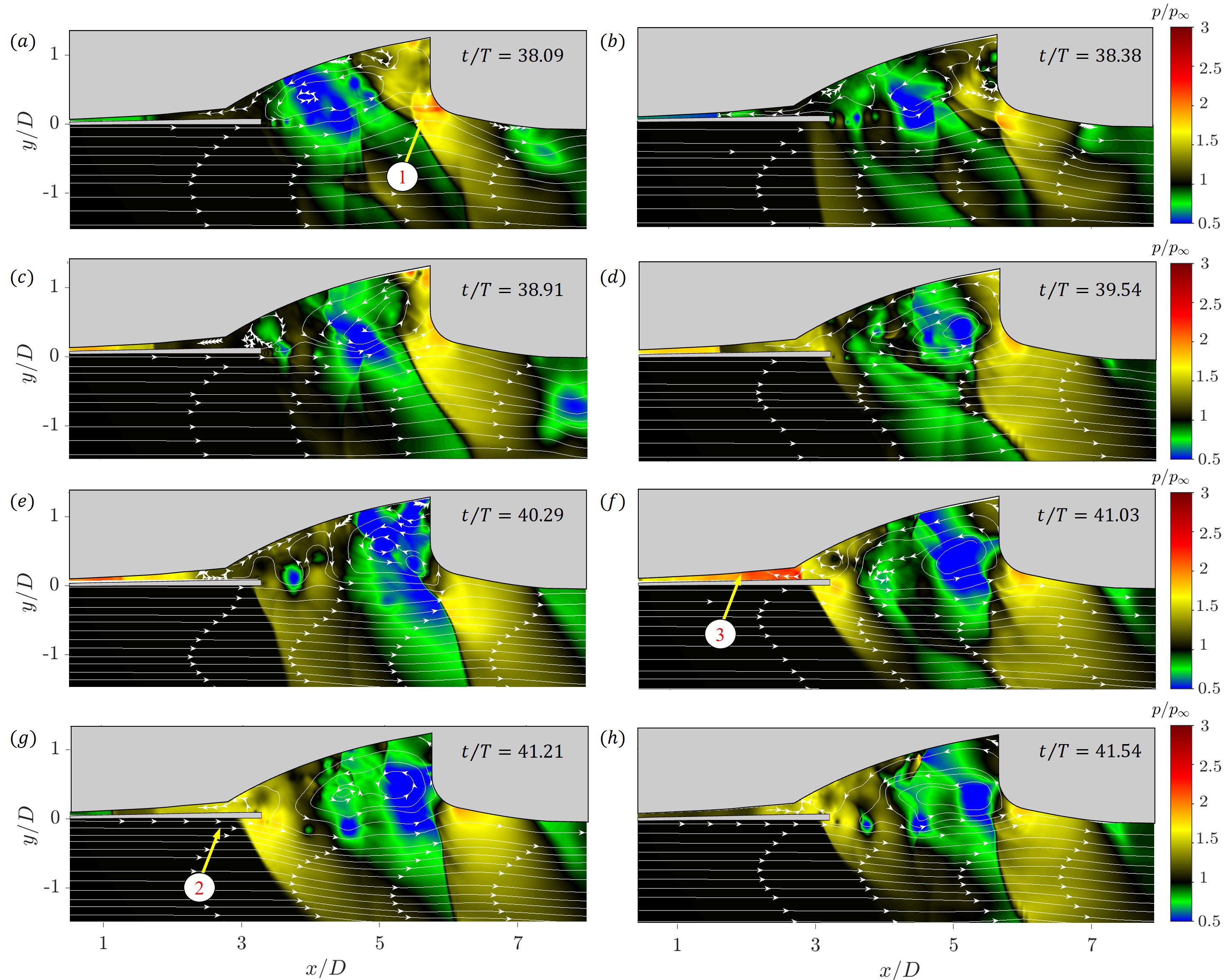}
   \caption{Instantaneous pressure contour plots at $M_\infty=1.2$ for control case C1 over one Rossiter oscillation cycle highlight key flow features: (1) High-pressure intensity concentrated at the aft wall, (2) Upstream propagating compression wave from leading edge of the cavity, and (3) Ejection of high-pressure flow from the sub-cavity. The superimposed streamlines further illustrate the dynamic exchange across the cavity opening, capturing both inflow and outflow processes, along with the formation of large recirculation zone within the cavity.  The non-dimensionalizing variables used here are $D=3.8$ m and $p_\infty=$ 2.6kPa.}
    \label{fig:passive_control_contour2}
\end{figure*}

\begin{figure*}
    \centering
   \includegraphics[width=1\linewidth] {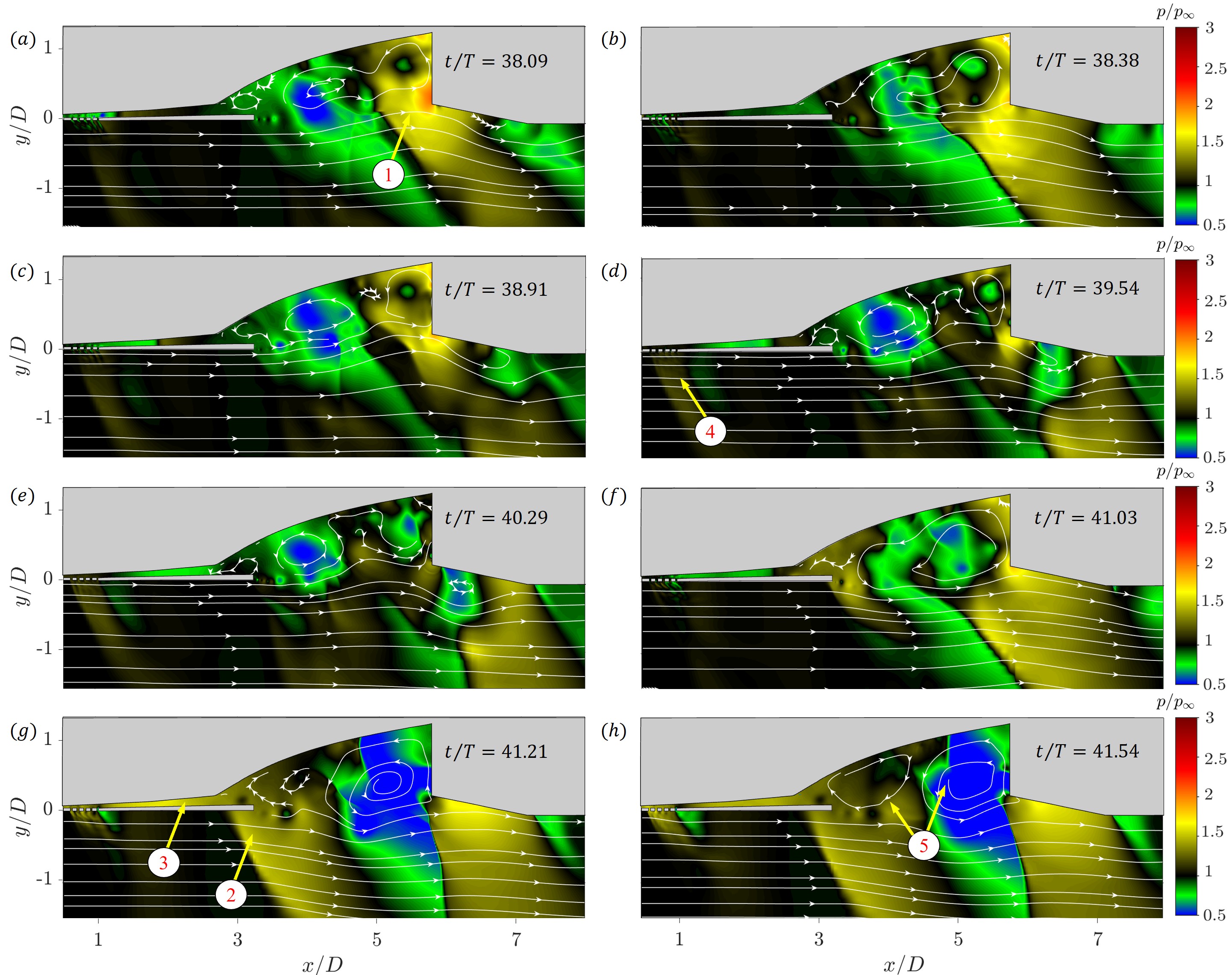}
    \caption{Instantaneous pressure contour plots at $M_\infty=1.2$ for control case C2 over one Rossiter oscillation cycle reveal the highlight key flow features: (1) High-pressure intensity concentrated at the aft wall, (2)  Upstream propagating compression from leading edge of the cavity, and (3) Ejection of low-pressure fluid flow from the sub-cavity, (4) Alternate compression and expansion regions over the sub-cavity vents, and (5) Two recirculation bubbles within the main cavity. The superimposed streamlines further illustrate the dynamic exchange across the cavity opening.  The non-dimensionalizing variables used here are $D=3.8$ m and $p_\infty=$ 2.6kPa.}
    \label{fig:passive_control_contour3}
\end{figure*}

\begin{figure*}
    \centering
    \includegraphics[width=0.8\linewidth] {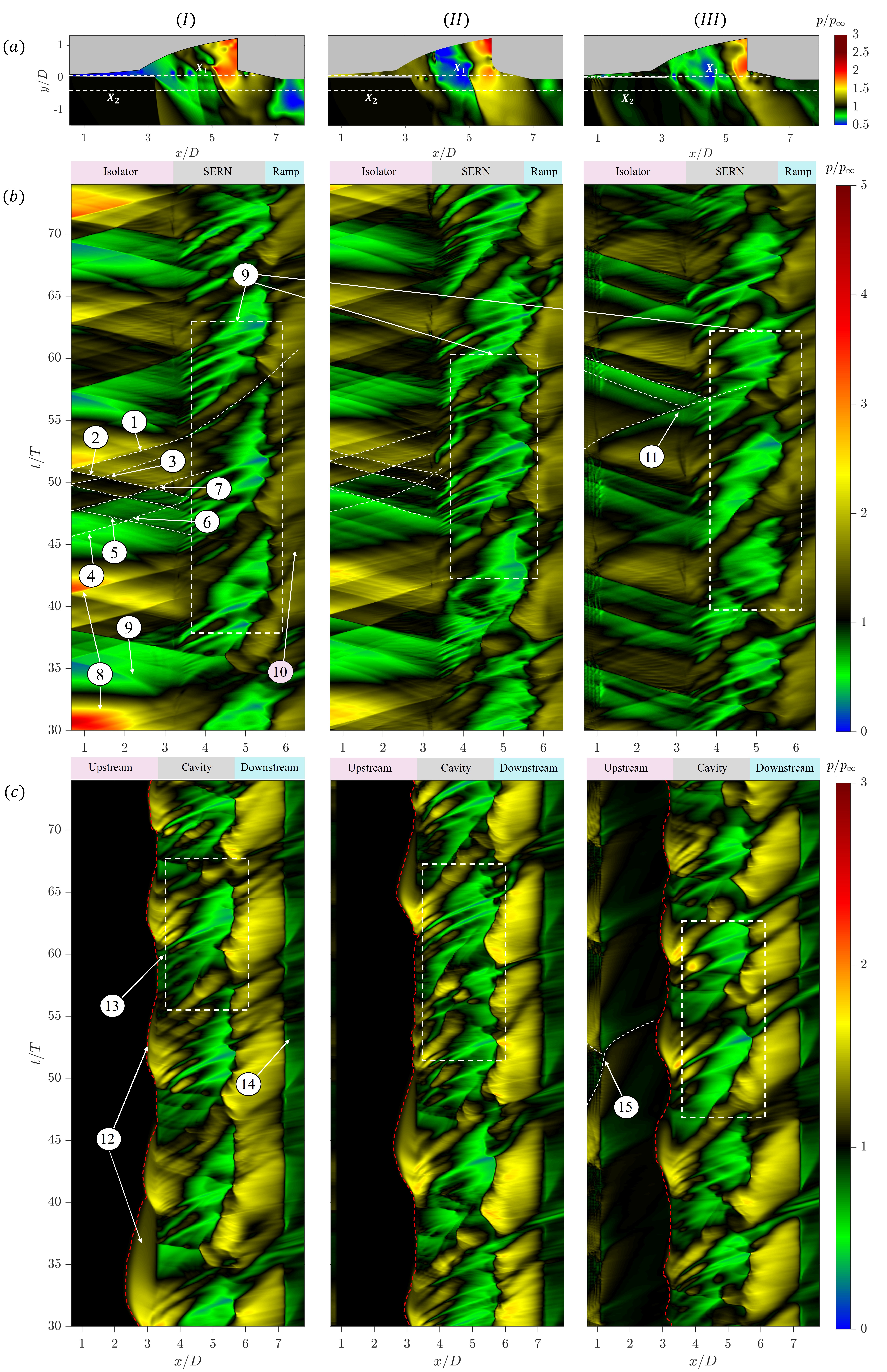}
    \caption{\href{https://youtube.com/shorts/w6E6eoQ5HBo?feature=share}{(Multimedia View)}(a) Instantaneous pressure contour plots at  $M_\infty=1.2$ for the geometries shown in Figure~\ref{fig:cavity_passive_control}, (b-c) time-averaged static pressure along line $X_1$, and line $X_2$ respectively, for (I) C0, (II) C1, and (III) C3. The key features highlight: (1) left-running compression wave (LW), (2) right-running compression wave (RW), (3) intersection of LW and RW compression waves, (4) left-running expansion wave, (5) right-running expansion wave, (6) intersection of LW and RW expansion waves, (7) intersection of RW compression wave and LW expansion wave, (8) high-pressure, (9) mixing zone, (10) downstream-propagating high-pressure disturbance, (11) Y-shaped wave, (12) upstream-propagating high-pressure disturbance, (13) upstream-propagating low-pressure disturbance, (14) low-pressure zone downstream, (15) upstream Y-shaped wave.  The non-dimensionalizing variables used here are $D=3.8$ m and $p_\infty=$ 2.6kPa.}
    \label{fig:passive_control_xt_contour}
\end{figure*}

\begin{figure*}
    \centering
   \includegraphics[width=1\linewidth] {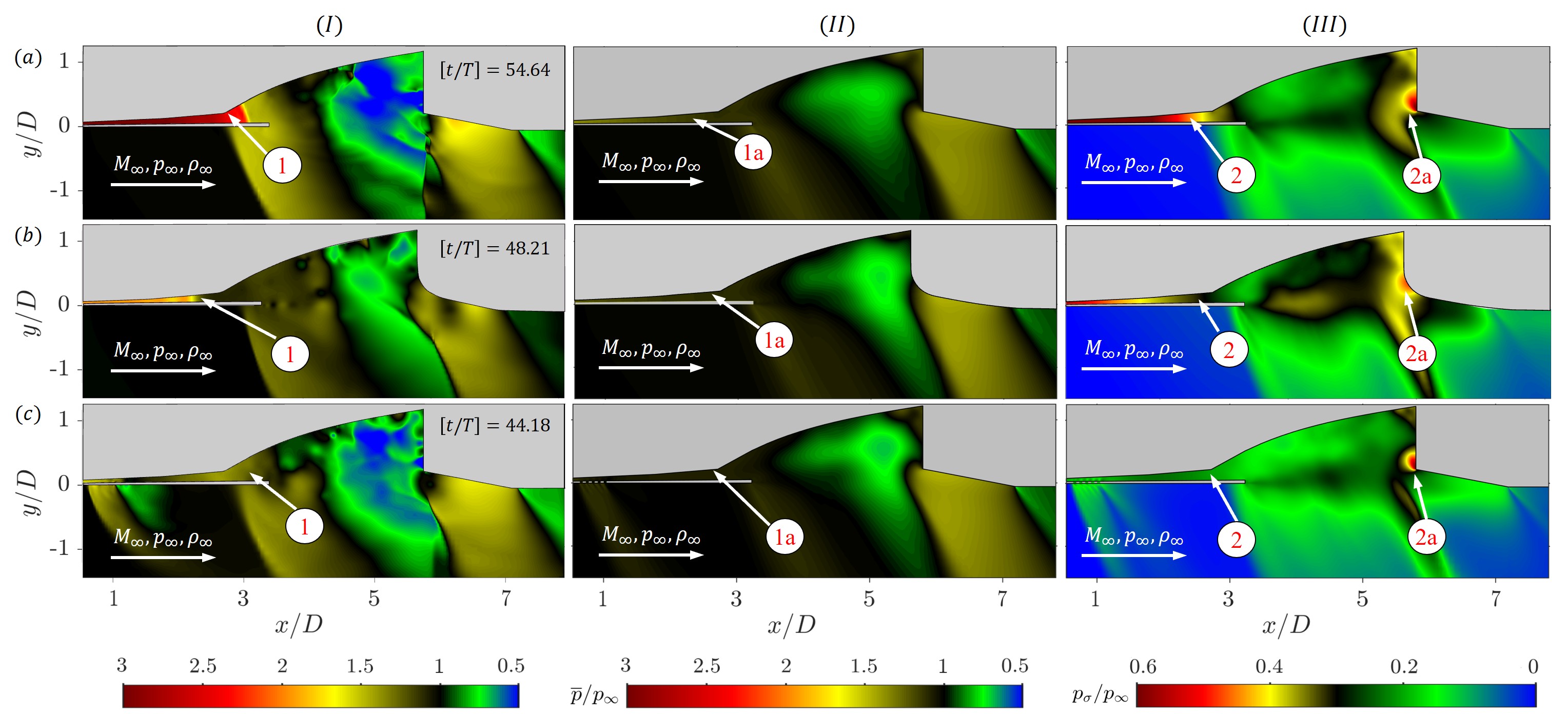}
    \caption{Non-dimensionalized contours for (I) Instantaneous pressure, (II) time-averaged pressure, and (III) standard deviation of pressure  for (a) C0, (b) C1, and (c) C2 geometries. The key features highlight: (1) High-pressure flow, (1a) Sub-cavity pressure intensity, (2) Sub-cavity pressure fluctuations, and (2a) Aft-wall pressure fluctuations. The non-dimensionalizing variables used here are $D=3.8$ m and $p_\infty=$ 2.6kPa.}
    \label{fig:passive_control_contour_all}
\end{figure*}

\begin{figure*}
    \centering
    \includegraphics[width=1\linewidth] {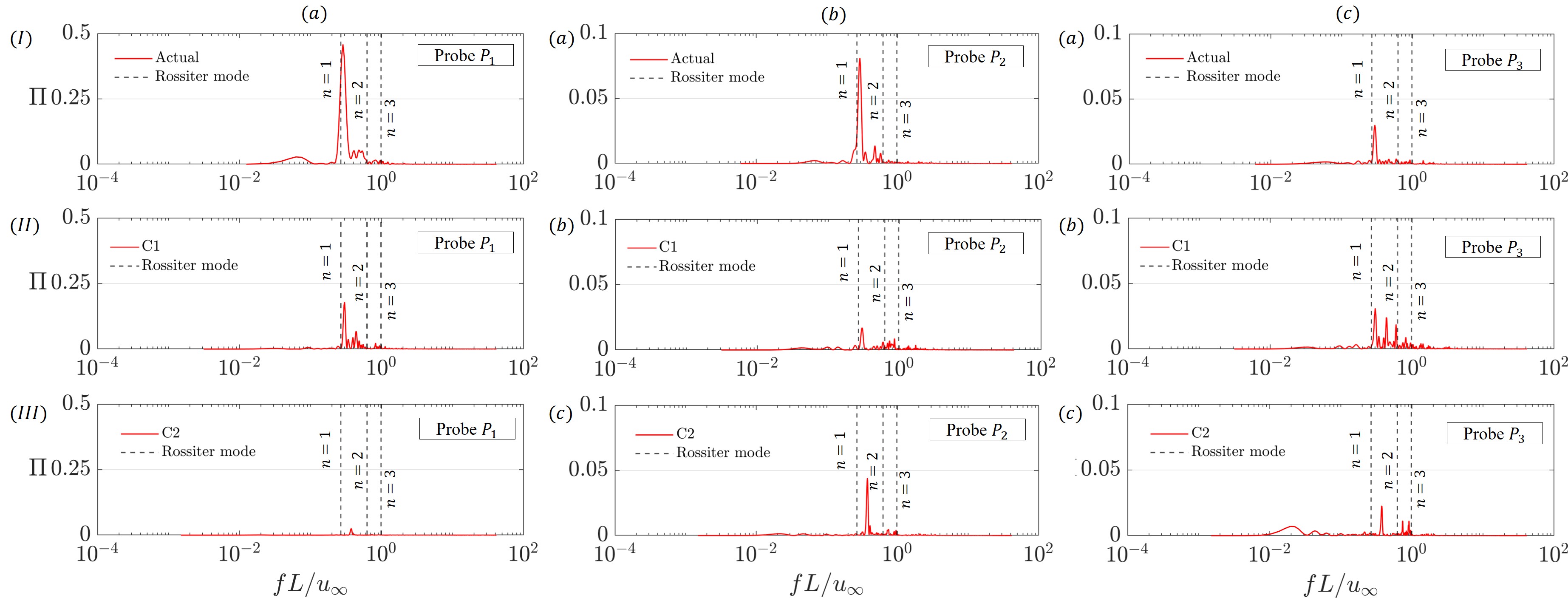}
    \caption{Non-dimensional power spectral density ($\Pi$) of static pressure fluctuations for the geometries shown in Figure~\ref{fig:cavity_passive_control} at three probe locations: (a) $P_1$ (sub-cavity end wall), (b) $P_2$ (trailing edge wall), and (c) $P_3$ (shear-layer midpoint), for (I) C0 (top row), (II) C1 (middle row), and (III) C2 (bottom row). The spectra are shown as a function of the non-dimensional frequency, $fL/u_\infty$. Dashed vertical lines indicate the first three Rossiter modes ($n=$1, 2, and 3). The non-dimensionalizing variables used here are $L=11.78$ m, $u_\infty = 354.08$ m/$s$, and $p_r=$ 1kPa.}
    \label{fig:passive_control_fft}
\end{figure*}

\begin{figure*}
    \centering
    \includegraphics[width=1\linewidth] {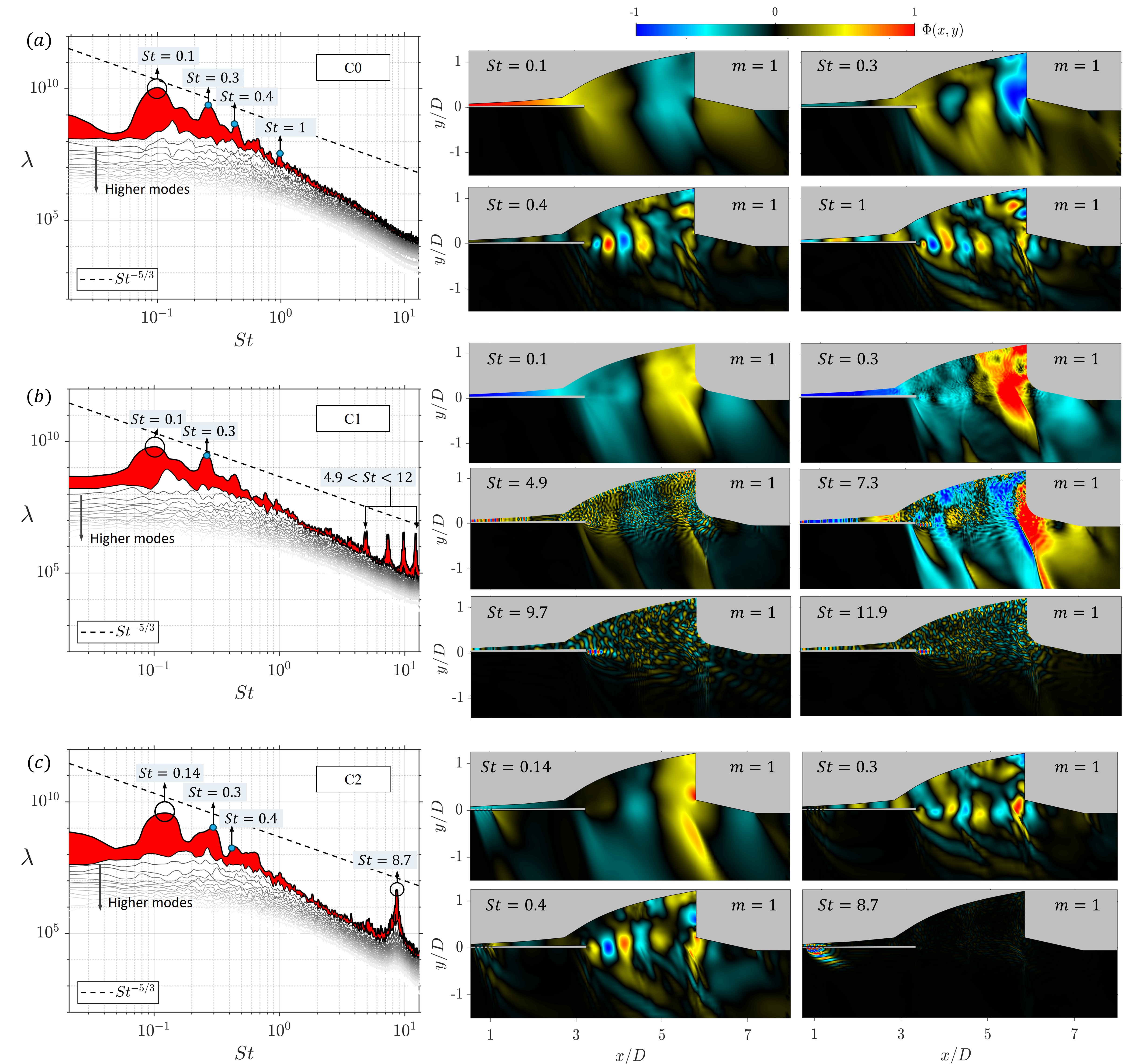}
    \caption{The SPOD energy spectra obtained from spatial pressure data, plotted across a range of non-dimensional frequencies $(St = fD/u_\infty)$, along with their corresponding dominant spatial modes, are presented for: (a) actual complex geometry C0, (b) C1 with a chamfered trailing edge, and (c) C2 with a ventilated sub-cavity. The red-shaded region represents the modal energy gap between the first and second modes. The black dashed line indicates the Kolmogorov power-law scaling $(St^{-5/3})$. The non-dimensionalizing variables used here are $D=3.8$ m and $u_\infty = 354.08$.}
    \label{fig:spod_analysis}
\end{figure*}

\begin{figure*}
    \centering
    \includegraphics[width=1\linewidth] {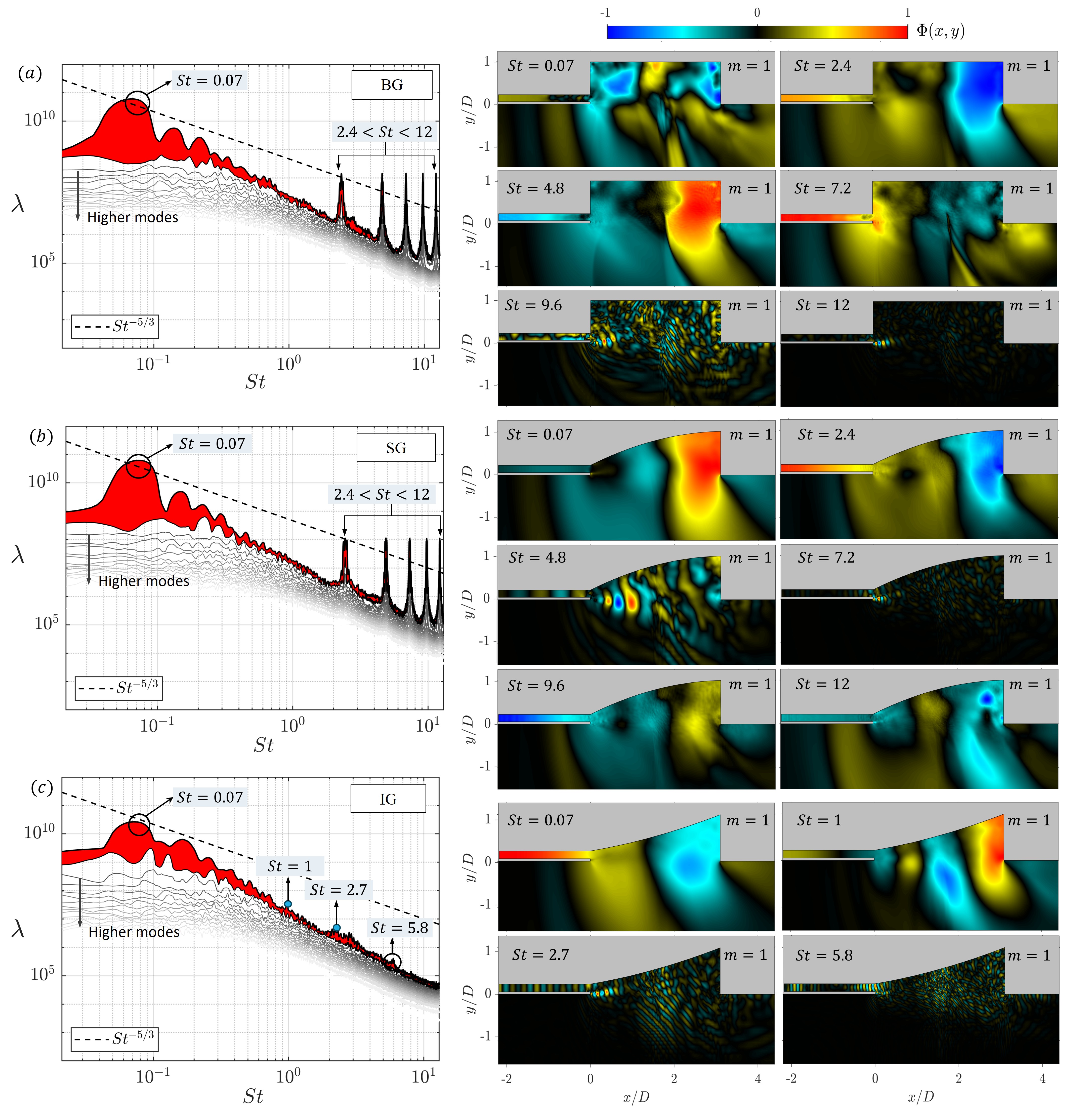}
    \caption{ The SPOD energy spectra obtained from spatial pressure data, plotted across a range of non-dimensional frequencies $(St = fD/u_\infty)$, along with their corresponding dominant spatial modes, are presented for different geometries: (a) baseline (BG), (b) SERN (SG), and (c) iSERN (IG).The red-shaded region represents the modal energy gap between the first and second modes. The black dashed line indicates the Kolmogorov power-law scaling $(St^{-5/3})$. The non-dimensionalizing variables used here are $D=3.8$ m and $u_\infty = 354.08$.}
    \label{fig:spod_baseline}
\end{figure*}

\begin{table*}
\centering
\caption{Static pressure statistics and dominant non-dimensional spectral characteristics at three probe locations for geometries in Figure~\ref{fig:schematic_shape} and Figure~\ref{fig:cavity_passive_control}, respectively. 
(a) Probe-I: sub-cavity end-wall ($x/D,y/D = -2.68,0.24$),
(b) Probe-II: aft-wall ($x/D,y/D = 0,0.24$),
(c) Probe-III: shear-layer midpoint ($x/D,y/D = 1.54,0.01$). `\% change' refers to variation relative to baseline (BG) and actual geometry (C0), respectively.
(Note: non-dimensionalized mean pressure - $\overline{p}/p_\infty$ ,
non-dimensionalized  standard deviation of pressure - $p_\sigma/p_\infty$,
non-dimensionalized frequency - $fL/u_\infty$,
and non-dimensionalized spectral power - $\Pi$).}
\begin{ruledtabular}
\begin{tabular}{lcc ccc ccc ccc}
\multicolumn{12}{c}{\textbf{Variation of cavity topology}} \\
\midrule

Geometry 
& 
& 
& \multicolumn{3}{c}{Probe-I}
& \multicolumn{3}{c}{Probe-II}
& \multicolumn{3}{c}{Probe-III} \\

\midrule

& $\overline{p}/p_\infty$ 
& $p_\sigma/p_\infty$
& $fL/u_\infty$ & $\Pi$ & \% change
& $fL/u_\infty$ & $\Pi$ & \% change
& $fL/u_\infty$ & $\Pi$ & \% change \\

\midrule

Baseline (BG)
& 1.2 & 0.6
& 0.07 & 1.2 & --
& 0.07 & 0.2 & --
& 0.07 & 0.03 & -- \\

SERN (SG)
& 1.5 & 0.9
& 0.08 & 2.3 & -60
& 0.08 & 0.4 & -75
& 0.38 & 0.08 & -- \\

iSERN (IG)
& 1.4 & 0.6
& 0.08 & 1.5 & -96
& 0.08 & 0.2 & -50
& 0.08 & 0.1 & -33 \\

\midrule
\midrule
\multicolumn{12}{c}{\textbf{Adapted control strategies}} \\
\midrule
Geometry 
& 
& 
& \multicolumn{3}{c}{Probe-I}
& \multicolumn{3}{c}{Probe-II}
& \multicolumn{3}{c}{Probe-III} \\

\midrule

& $\overline{p}/p_\infty$ 
& $p_\sigma/p_\infty$
& $fL/u_\infty$ & $\Pi$ & \% change
& $fL/u_\infty$ & $\Pi$ & \% change
& $fL/u_\infty$ & $\Pi$ & \% change \\

\midrule

Actual (C0)
& 1.42 & 0.67
& 0.28 & 0.45 & --
& 0.28 & 0.08 & --
& 0.30 & 0.03 & -- \\

Control-1 (C1)
& 1.34 & 0.55
& 0.30 & 0.18 & -60
& 0.30 & 0.02 & -75
& 0.30 & 0.03 & -- \\

Control-2 (C2)
& 1.08 & 0.20
& 0.37 & 0.02 & -96
& 0.37 & 0.04 & -50
& 0.37 & 0.02 & -33 \\

\end{tabular}
\end{ruledtabular}
\label{tab:pressure_freq_psd}
\end{table*}

\begin{table}[htbp]
\centering
\caption{Mass flux ratios for actual geometry and control cases. The parameter $\eta_\text{in} = {\dot{m}_i}/{\dot{m}_a}$  denotes the ratio of mass entering the cavity to the mass entering the cavity for the actual geometry, while $\eta_\text{vent} = {\dot{m}_v}/{\dot{m}_a}$ represents the ratio of ventilated mass through the slot to the mass entering the cavity for the actual geometry.}
\begin{ruledtabular}
\begin{tabular}{lcc}
Geometry 
& $\eta_\text{in}$ 
& $\eta_\text{vent}$ \\ \midrule

Actual        
& 1.00 & -- \\

C1             
& 1.06 & -- \\

C2             
& 0.62 & 0.06 \\

\end{tabular}
\end{ruledtabular}
\label{tab:mass_stats}
\end{table}

Table~\ref{tab:mass_stats} provides the mass flux statistics observed along the cavity opening (line AB) for the three geometries. The BG experiences highest mean mass flux and IG experiences 32\% less mean flow compared to BG. The mass fluctuations observed also follow the same trend with BG experiencing 21\% higher fluctuations than IG. The ratio of mass flux enetring to mass exiting along line AB, given as $\eta$ ($\eta = \dot{m_{i}}/\dot{m_{e}}$), is also least in IG. The skewness parameter $\beta_1$ describe how the distribution of instantaneous values deviates from a symmetric (Gaussian) distribution. IG exhibits the value closest to zero, suggesting a nearly symmetric distribution of mass-flux fluctuations. $\beta_2$ quantifies the intermittency and extremeness of events. IG thereby showcases the highest intermittency. Therefore, IG  promotes a more confined flow environment, in which mass exchange across the cavity opening is suppressed on average but dominated by sporadic, high-intensity entrainment events. The variation in topology also results in significant difference in the flow features observed during the mass flow/shear layer deflection cycles. 

During the inward deflection, the sub-cavity region experiences negative streamwise velocity along its length in BG and SG. In contrast, SG develops a largely stagnant flow within the sub-cavity region. The leading edge expansion fan (feature `2’) is considerably weaker in IG due to less available volume to expand. The sub-cavity region in IG also lacks the distinct low-pressure region (feature `3’) observed in BG and SG. This results in low pressure differential across the primary cavity and sub-cavity, thereby reducing mass inflow into the sub-cavity and thus unsteady pressure loads on the sub-cavity walls are also reduced. The KH-instability present in the shear layre is more pronounced in IG. The shear-layer roll-up and subsequent breakdown into smaller vortical structures (feature `4’) occur more rapidly in IG. The circulatory vortex inside the cavity also weakens progressively from the BG to IG. The reflected compression wave (feature `5’) on the trailing edge wall appears in all geometries and the intensity significantly varies with BG showcasing higher intensity.

During outward deflection of shear layer, IG experiences larger supersonic flow regions within the primary cavity (feature `6’). The volume occupied by the recirculation bubble (feature `7') is also reduced in IG. Along with that IG does not exhibit supersonic inflow at the sub-cavity neck unlike the other two geometries. The high-pressure flow (feature `10’) exits the sub-cavity at higher intesnity in SG compared to other two cases. The contours shown in Figure~\ref{fig:cavity_shape_shear_mean} also show that SG experiences higher pressure fluctuations  (see Figure~\ref{fig:cavity_shape_shear_mean}, feature `3') in sub-cavity region. IG shows less fluctuations on sub-cavity region and trailing edge wall in compariosn with other two cases. Furthermore, the low-pressure region ( feature `2') confined within the primary cavity is the smallest in IG thereby leading to less fluid inflow during shear layer oscillations. On prominent thing observed is upstream disturbance propagation is greatest in BG and the smallest in IG. The influence on upstream flow is the least in IG.

Figure~\ref{fig:cavity_shape_fft2} shows the power spectra of the pressure distribution at three probe locations, the sub-cavity end wall ($P_1$), trailing edge wall ($P_1$), and shear layer (mixing layer) midpoint ($P_3$). Table \ref{tab:pressure_freq_psd} provides the pressure statistics observed at these probe points. SG experiences the higheset mean pressure ($\overline{p}/p_{\infty}$) followed by IG. SG also experiences highest standard deviation in pressure ($p_{\sigma}/p_{\infty} = 0.93$). This was evident in Figure~\ref{fig:cavity_shape_shear_mean}. BG pressure values are in an intermediate range between SG and IG. The peak dominant frequency observed for SG and IG shift by 14\% in comparison with BG. Across probes $P_1$ and $P_2$, the peak spectra pressure power ($\Pi$) is 91\% higher than BG, followed by IG by 25\%. Whereas, at probe $P_3$ , the trend is reversed. IG showcases higher spectra power by 90\% compared to BG. This indicates that IG showcases shear-layer dominant behavior.

\subsection{Passive control in complex cavity-sub-cavity system}
Many control techniques have been tested to reduce cavity acoustic tones. Both active~\cite{colonius1999} and passive control~\cite{turpin2020,Das2017,seker2023,roshko1955,perng2001} have been used for this purpose. These approaches have been applied in several flow studies beyond cavity flows, including shock-induced transonic buffet~\cite{crouch1997}, shock wave/laminar boundary layer interaction~\cite{robinet2007}, flow around a swept parabolic body~\cite{mack2008}, and axisymmetric wake flows~\cite{meliga2010}. Passive control techniques are the easiest to implement. Considering our complex cavity-sub-cavity system is derived from a scramjet, which features no moving parts, passive control methods are more suitable. Widely tested passive controls include the use of spoilers, perforated walls, and modifications to the cavity's leading and/or trailing edge. These concepts have proved to be very effective in reducing energetic tones. 

Two passive control methods were investigated, referred to as C1 and C2. In C1 the trailing edge wall of the cavity is modified by providing a chamfered curvature. In C2, ventilation slots were introduced in the sub-cavity region. The detailed schematics of C1 and C2 are presented in Figure~\ref{fig:cavity_passive_control}. The comparisons are made in terms of pressure values at the control points obtained from the sub-cavity end wall ($P_1$), trailing edge wall ($P_2$) and shear layer midpoint ($P_3$), which are the locations where the pressure fluctuations are observed to be maximum.  The spectrum obtained from FFT analysis of the static pressure measurements are also presented for the actual case and the control cases for comparison purposes. Pressure contours with streamlines for various time steps are used for comparison as well (see Figure~\ref{fig:passive_control_contour1}-\ref{fig:passive_control_contour3}). 

The temporal evolution of pressure along line $X_1$ (extending from $[x/D, y/D] = [0.54, 0.25]$ to $[6.4, 0.25]$ inside the cavity) and line $X_2$ (from $[0.54, -0.4]$ to $[7.6, -0.4]$ in the external flow) is presented in Fig.~\ref{fig:passive_control_xt_contour}. The $x$–$t$ contours clearly depict the propagation and interaction of pressure waves governing the cavity dynamics. Along $X_1$ for the C0 configuration, distinct left-running (LW, feature ‘1’) and right-running (RW, feature ‘2’) compression waves are observed, whose opposite slopes in the $x$–$t$ plane indicate bidirectional propagation. Their intersection (feature ‘3’) results in localized amplification of pressure fluctuations. Similarly, left- and right-running expansion waves (features ‘4’ and ‘5’) are identified, and their coincidence (feature ‘6’) produces low-pressure disturbances that subsequently convect downstream. Interactions between LW compression waves and RW expansion waves further generate coherent disturbances that travel downstream over time. Periodic high-pressure bursts (feature ‘8’) are evident, corresponding to shear-layer impingement and associated shock oscillations. The region $x/D = 3.7$–6 (boxed area) marks the dominant mixing zone (feature ‘9’), where repeated shock–shear layer interactions produce strong nonlinear wave activity. Comparable features are observed in C1, though with slightly attenuated intensity. In contrast, C2 exhibits weaker LW and RW signatures and a more discrete, less chaotic mixing region, indicating suppression of wave amplification. Along $X_2$, the contours illustrate the upstream influence of cavity oscillations. The upstream-propagating pressure disturbances, highlighted by the red dotted trajectory, demonstrate the acoustic feedback mechanism. Alternating high- and low-pressure bands (features ‘12’ and ‘13’) correspond to periodic shear-layer deflection and shock motion. In C2, ventilation on the sub-cavity wall leads to the emergence of a distinct Y-shaped wave structure (feature ‘15’) significantly altering the upstream flow.

Figure~\ref{fig:passive_control_contour_all} shows the instantaneous, time-averaged, and standard deviation of pressure over the domain. As clearly observed from the contours, the C1 slightly reduces the pressure intensity observed in the sub-cavity region and the trailing end wall. While C2 effectively reduces the pressure intensity in the sub-cavity region, the trailing edge still experiences high-intensity pressure. C2 actively suppresses the pressure fluctuations in the sub-cavity and trailing end-wall region, unlike C1, where it induces stronger disturbances in the shear layer (mixing layer), and the trailing end wall experiences comparatively higher pressure fluctuations than the actual case.

The power spectra obtained at the control points are shown in Figure~\ref{fig:passive_control_fft}. The graphs show the peaks obtained for the control cases in relation to Rossiter frequency. The C1 peaks on all control points alligns with the Rossiter second mode frequency, indicating the flow is governed by the feedback aeroacoustic mechanism. While in C2, the dominant spectral peaks deviate significantly from the Rossiter modes at low frequencies, while secondary peaks at higher frequencies exhibit alignment with the higher Rossiter modes. This behaviour indicates a shift in the governing dynamics from a purely acoustically driven shear-layer resonance at low frequencies, with the aeroacoustic feedback mechanism becoming active only at higher frequencies.

Table~\ref{tab:mass_stats} provides the mass flux ratios for the actual and control geometries. $\eta_{in}$ is the ratio between mass entering the control case through cavity opening (line AB) to mass entering the actual geometry.
C1 exhibits 6\% higher inflow in the cavity opening compared to the actual geometry. Whereas, C2 shows 38\% reduction in the inflow. Additionally, $\eta_{vent}$ is calculated which is the ratio between mass flow entering through the ventilated slots to the mass entering the actual geometry across line AB. Approximately 6\% of the inflow is diverted through the ventilation slot. 

Table~\ref{tab:pressure_freq_psd} provides the pressure statistics at the control points for the actual and control cases. The C2 effectively dampens the peak pressure spectral power ($\Pi$) at the sub-cavity end wall by 96\%, while C1 reduces it by 60\%. However, C2 now experiences higher loading at the trailing end wall compared to C1, which results in 75\% reduction in peak spectral power. C2 significantly reduces the fluctuations at the control point in the shear layer (mixing layer) by 33\%, whereas no significant changes are observed for C1 at this point. Thus, providing ventilation proves to be effective in reducing the pressure fluctuations by breaking the periodicity of the flow, while the periodic nature of the flow in C1 remains unchanged, the magnitude of the fluctuations slightly decreases at the control points.

\subsection{Spectral Proper Orthogonal Decomposition (SPOD) analysis}
Spectral Proper Orthogonal Decomposition (SPOD) extracts the main spectral features of a flow field, optimally capturing the two-point space–time correlations, thus providing modes that evolve coherently in space and time\cite{schmidt2018,schmidt2022,towne2018}. Following the works of Towne et al.\cite{towne2018} and Schmidt and Colonius\cite{schmidt2020}, the standard SPOD algorithm has been employed. For a detailed mathematical derivation and algorithmic implementation of SPOD, the reader is referred to \cite{towne2018} and Schmidt and Colonius\cite{schmidt2020}. Here, we briefly summarize the notation adopted in this study. 

We denote by
\begin{equation}
\mathbf{q}_i = \mathbf{q}(t_i), \qquad i = 1, 2, \ldots, n_t,
\end{equation}
the ensemble of $n_t$ snapshots of the statistically stationary flow field, with the mean removed. Following Welch’s method, the data are divided into $n_{\mathrm{blk}}$ overlapping blocks, each containing $n_{\mathrm{fft}}$ snapshots. If the blocks overlap by $n_{\mathrm{ovlp}}$ snapshots, the $j$-th column in the $k$-th block is given by
\begin{equation}
\mathbf{q}_j^{(k)} = \mathbf{q}_{j+(k-1)(n_{\mathrm{fft}}-n_{\mathrm{ovlp}})+1},
\end{equation}
where $k = 1,2,\ldots,n_{\mathrm{blk}}$ and $j = 1,2,\ldots,n_{\mathrm{fft}}$.

A windowed temporal discrete Fourier transform is then applied, and the Fourier realizations at the $l$-th frequency, $\hat{\mathbf{q}}_l^{(j)}$, are assembled into the matrix
\begin{equation}
\hat{\mathbf{Q}}_l =
\left[
\hat{\mathbf{q}}_l^{(1)},
\hat{\mathbf{q}}_l^{(2)},
\ldots,
\hat{\mathbf{q}}_l^{(n_{\mathrm{blk}})}
\right].
\end{equation}

The SPOD modes $\boldsymbol{\Phi}_l$ and the corresponding eigenvalues $\boldsymbol{\Lambda}_l$ are obtained as the eigenpairs of the weighted CSD matrix
\begin{equation}
\mathbf{S}_l
= \frac{1}{n_{\mathrm{blk}}}
\hat{\mathbf{Q}}_l
\hat{\mathbf{Q}}_l^{*}
\mathbf{W},
\end{equation}
where $\mathbf{W}$ is a positive-definite Hermitian matrix accounting for component-wise weighting and numerical quadrature. When the number of spatial degrees of freedom exceeds the number of snapshots, the equivalent reduced eigenvalue problem is solved:
\begin{equation}
\frac{1}{n_{\mathrm{blk}}}
\hat{\mathbf{Q}}_l^{*}
\mathbf{W}
\hat{\mathbf{Q}}_l
\boldsymbol{\Psi}_l
=
\boldsymbol{\Psi}_l
\boldsymbol{\Lambda}_l,
\end{equation}
where $\boldsymbol{\Psi}_l$ contains the (unscaled) expansion coefficients. The SPOD modes at the $l$-th frequency are finally recovered as
\begin{equation}
\boldsymbol{\Phi}_l
=
\frac{1}{\sqrt{n_{\mathrm{blk}}}}
\hat{\mathbf{Q}}_l
\boldsymbol{\Psi}_l
\boldsymbol{\Lambda}_l^{-1/2}.
\end{equation}

The snapshots have been sampled with a dimensionless time equal to 0.012, and a total of $N_t$ = 7500 snapshots were considered. For the SPOD analysis, $N_f$ = 1024 snapshots have been used, considering a Hamming temporal window and a 50\% overlap among blocks. This resulted in $N_b$ = 13 blocks. Figure~ \ref{fig:spod_analysis}-\ref{fig:spod_baseline} shows the energy spectra as a function of non-dimensional frequency ($St = fD/u_\infty$) along with their spatial modes for the cases considered in Section V.

For the control cases considered in Section \ref{sec: results}d, the most prominent modal separation between the eigen values associated with the first and the second SPOD modes is observed at  $St \approx 0.1$ for all the cases. This refers to as low-rank behavior and often indicates that a physically dominating hydrodynamic mechanism exists in the flow field. The energy spectra for the C1 with a chamfered trailing edge shows multiple dominant peaks at higher frequencies ($St \geq 4$). The spatial modes show compact wave packet structures, indicating that the Kelvin–Helmholtz–type (KH) spatial instability is dominating at higher frequency ranges. Whereas large Orr-type/Bluff body like structures dominate at lower frequencies ($St \leq 0.1$), indicating accoustic dominant behaviorr. 

The energy spectra for C2 with a ventilated sub-cavity exhibits a single dominant peak at $St = 8.7$. The spatial modes at $St = 8.7$ exist as wavepacket structures and are confined near the slot region. As a reference for the spectral behavior, the blue dashed line in Figure \ref{fig:spod_analysis} represents the Kolmogorov power law ($St^{-5/3}$), which is a well-established representation of the energy spectrum in turbulent flows and serves as a benchmark for assessing the spectral characteristics of the analyzed data. The SPOD energy spectrum curve aligns with this well-known benchmark. As the power law scaling is not valid at low frequencies, where relatively larger coherent structures characterize the flow is evident in Figure~ \ref{fig:spod_analysis} and \ref{fig:spod_baseline}. It is noted that at higher frequencies ($St \geq 0.4$), Kelvin–Helmholtz-type (KH) spatial instability is dominant in the flow, whereas at lower frequencies ($St \leq 0.1$) large-scale bluff structures dominate the flow field.

For geometries considered in  \ref{sec: results}c, the dominant mode is observed at $St = 0.07$. For the BG, five distinct dominant peaks emerge at higher frequencies ($2.4 \leq St \leq 12$). Figure~ \ref{fig:spod_baseline} shows the spatial modes at these frequencies. At $St < 0.1$, large Orr-type structures predominate the flow, with the mode being more distinctly defined. At higher frequencies ($2.4 \leq St \leq 12$), it is observed that these structures exhibit both large Orr-type and lobe-like formations, dominated by cavity resonance and wave packet structures indicative of Kelvin-Helmholtz (KH) instability. Notably, at $St = 3$, spatial modes display wave packet-like structures near the cavity's leading edge, suggesting that the flow resonance is primarily influenced by shear-layer oscillations. At frequencies $St > 3$, similar Orr-type structures continue to dominate the flow, indicating acoustic dominant behaviour.

In SG, a similar trend is observed, with five dominant peaks at higher frequencies ($2.4\leq St\leq12$). At $St=0.07$, large structures dominate the flow. Unlike BG, at $St = 0.1$, mode 1 reveals large wave packets emerging from the leading edge point, indicating that shear is exhibiting dominance at this frequency. This is more pronounced for $St = 2.4$ and $St= 3$, where more defined wave packet-like structures dominate. For $St>3$, large lobe structures are found to be dominant, suggesting that for $2.4\leq St\leq 3$, KH instability is exhibited, while at higher frequencies, acoustic modes dominate for SG. In contrast, IG exhibits a dominant mode at $St = 0.07$ and a minor peak at a frequency of $St = 5.8$. At the lower frequency of $St = 0.07$, the spatial modes reveal large Orr-type structures. For $St > 0.1$, a combination of small and large-scale wavepackets emerges from the leading edge, indicating that $St > 0.1$ is also dominated by KH-instability.

\section{Conclusion}
\label{sec:conclusion}
In summary, this research provides new insights into the unsteady flow dynamics and pressure loading characteristics of the complex cavity-sub-cavity system. Through a combination of experimental validation and high-fidelity Detached Eddy Simulation, the study demonstrates that cavity topology and Mach number critically influence flow unsteadiness and resonance phenomena. It was identified that as Mach number increases from subsonic to supersonic, the pressure load and peak spectral power acting on the sub-cavity end-wall monotonically increase. Two passive control strategies were explored to mitigate the pressure oscillations observed in the cavity-sub-cavity system. The control configuration with a chamfered aft-wall (C1) exhibits a 60\% reduction in peak spectral power. In contrast, the configuration with a ventilated sub-cavity (C2) proves to be the most effective, achieving a 96\% suppression of pressure fluctuations while also stabilizing the shear-layer. These observations are consistently supported by both spectral and SPOD analyses. The SPOD results further reveal a restructuring of dominant modal flow features, along with the emergence of new modal peaks at higher frequencies for the control cases. These findings underscore the significance of tailored geometric modifications in mitigating adverse pressure oscillations in practical aerospace applications.

\begin{acknowledgments}
The authors gratefully acknowledge the PARAMSEVA clusters for providing computational resources. Financial support from DRDL–DRDO Hyderabad, the Japan Science and Technology (JST) Fellowship, and the JICA 2.0 Friendship Grants is also sincerely acknowledged.
\end{acknowledgments}

\section*{References}
\bibliography{[1]_references}
\onecolumngrid
\PRLsep
\end{document}